\documentclass[12pt,tightenlines]{revtex4-1} 
\usepackage{color}
\usepackage{graphicx}
\usepackage[small,bf]{caption}
\usepackage{epsfig,wrapfig,sidecap}
\usepackage{fancyhdr,lastpage}
\usepackage[T1]{fontenc}
\usepackage[latin1]{inputenc}
\usepackage[english]{babel}
\usepackage{helvet}

\newcommand{\shorttitle}{HEP-FCE Working Group Reports}
\newcommand{\projectnum}{}
\newcommand{\piname}{}

\newcommand{\smarial}[1]{\fontsize{10pt}{0pt} \em{#1}}

\begin{document}


\noindent\line(1,0){470}\\

\vspace{0.5cm}

\noindent{\bf \scshape{{\centerline {High Energy Physics Forum for Computational
    Excellence:}}\\ 
{\centerline{Working Group Reports}}}}

\bigskip

\noindent{\bf \scshape{{\centerline {I. Applications Software}}}}\\
\noindent{\bf \scshape{{\centerline {II. Software Libraries and Tools}}}}\\
\noindent{\bf \scshape{{\centerline {III. Systems}}}}

\vspace{1.5cm}

\noindent{\bf Lead Editors:} Salman Habib$^1$ and Robert Roser$^2$
(HEP-FCE Co-Directors)\\

\noindent{\bf Applications Software Leads:} Tom LeCompte$^1$, Zach
Marshall$^3$\\ 
\noindent{\bf Software Libraries and Tools Leads:} Anders
Borgland$^4$, Brett Viren$^5$\\ 
\noindent{\bf Systems Lead:} Peter Nugent$^3$

\vspace{1.5cm}

\noindent{\bf Applications Software Team:}\\
Makoto Asai$^4$, Lothar Bauerdick$^2$, Hal Finkel$^1$,
Steve~Gottlieb$^6$, Stefan Hoeche$^4$,\\ Tom LeCompte$^1$, Zach
Marshall$^3$, Paul Sheldon$^7$, Jean-Luc Vay$^3$

\bigskip

\noindent{\bf Software Libraries and Tools Team:}\\
Anders Borgland$^4$, Peter Elmer$^8$, Michael Kirby$^2$, Simon
Patton$^3$, Maxim Potekhin$^3$,\\ Brett Viren$^3$, Brian Yanny$^2$

\bigskip

\noindent{\bf Systems Team:}\\
Paolo Calafiura$^3$, Eli Dart$^3$, Oliver Gutsche$^2$, Taku
Izubuchi$^5$, Adam Lyon$^2$,\\ Peter Nugent$^3$, Don Petravick$^9$

\vspace{1.5cm}

{\em
\noindent$^1$Argonne National Laboratory, 9700 S. Cass Ave., Lemont,
IL 60439\\
\noindent$^2$Fermi National Accelerator Laboratory, Batavia, IL 60510\\
\noindent$^3$Lawrence Berkeley National Laboratory, 1 Cyclotron Road,
Berkeley, CA 94720\\
\noindent$^4$SLAC National Accelerator Laboratory, 2575 Sand Hill
Road, Menlo Park, CA 94025\\
\noindent$^5$Brookhaven National Laboratory, Upton, NY 11973\\
\noindent$^6$Department of Physics-SW117, Indiana University,
Bloomington, IN 47405,\\ 
\noindent$^7$Department of Physics and Astronomy, Vanderbilt
University, TN 37235\\ 
\noindent$^8$Department of Physics, Jadwin Hall, Princeton
University, NJ 08544\\ 
\noindent$^9$National Center for Supercomputing Applications, UIUC,
Urbana, IL 61801\\  
}

\newpage

\begin{center}
{\bf\scshape{Abstract}}
\end{center}

Computing plays an essential role in all aspects of high energy
physics. As computational technology evolves rapidly in new
directions, and data throughput and volume continue to follow a steep
trend-line, it is important for the HEP community to develop an
effective response to a series of expected challenges. The computing challenges
require adopting new strategies in algorithms, software, and hardware
at multiple levels in the HEP computational pyramid. A significant
issue is the human element -- the need for training a scientific and technical
workforce that can make optimum use of state-of-the-art computational
technologies and be ready to adapt as the landscape changes.

In order to help shape the desired response, the HEP Forum for
Computational Excellence (HEP-FCE) initiated a roadmap planning
activity with two key overlapping drivers -- 1) software
effectiveness, and 2) infrastructure and expertise advancement. These
drivers had been identified in a number of previous studies, including
the 2013 HEP Topical Panel on Computing, the 2013 Snowmass Study, and
the 2014 P5 report. The HEP-FCE formed three working groups, 1)
Applications Software, 2) Software Libraries and Tools, and 3) Systems
(including systems software), to provide an overview of the current
status of HEP computing and to present findings and opportunities for
the desired HEP computational roadmap. A choice was made to focus on
offline computing in HEP experiments, even though there can be
nontrivial connections between offline and online computing.

This document begins with a summary of the main conclusions and
directions contained in the three reports, as well as a statement of
the cross-cutting themes that emerge from them. Because the scope of
HEP computing is so wide, it was impossible to give every technical
area its due in the necessarily finite space of the individual
reports. By covering some computational activities in more detail than
others, the aim has been to convey the key points that are independent
of the individual research projects or science directions. The three
main reports follow in order after the summary.

The Applications Software Working Group undertook a survey of members
of the HEP community to ensure a broad perspective in the
report. Albeit not a complete sample of the HEP community, the
respondents covered a range of experiments and projects. Several
dozens of applications were discussed in the responses. This mass of
information helped to identify some of the current strengths and
weaknesses of the HEP computing effort.

A number of conclusions have emerged from the reports. These include
assessments of the current software base, consolidation and management
of software packages, sharing of libraries and tools, reactions to
hardware evolution (including storage and networks), and possibilities
of exploiting new computational resources. The important role of
schools and training programs in increasing awareness of modern
software practices and computational architectures was emphasized. A
thread running across the reports relates to the difficulties in
establishing rewarding career paths for HEP computational
scientists. Given the scale of modern software development, it is
important to recognize a significant community-level software
commitment as a technical undertaking that is on par with major
detector R\&D.

Conclusions from the reports have ramifications for how computational
activities are carried out across all of HEP. A subset of the
conclusions have helped identify initial actionable items for HEP-FCE
activities, with the goal of producing tangible results in finite time
to benefit large fractions of the HEP community. These include
applications of next-generation architectures, use of HPC resources
for HEP experiments, data-intensive computing (virtualization and
containers), and easy-to-use production-level wide area networking. A
significant fraction of this work involves collaboration with DOE ASCR
facilities and staff.

\newpage

\noindent
{\bf \scshape Table of Contents}

\begin{tabular}{lr}

\noindent {\scshape {HEP-FCE Working Group Reports
    Summary}}...............................................................& 1\\
\\
\noindent {\scshape {1~Applications Software Report:
    Introduction}}........................................................& 4\\
\noindent {\scshape {2~Packages in Common Use}}................................................................................................& 4\\
\hspace{0.7cm}{\em 2.1~Experiment and Phenomenology}...................................................................................& 4\\
\hspace{0.7cm}{\em 2.2~Accelerator Physics}.......................................................................................................& 5\\
\hspace{0.7cm}{\em 2.3~Lattice QCD}.................................................................................................................& 6\\
\hspace{0.7cm}{\em 2.4~Computational Cosmology}............................................................................................& 7\\
\hspace{0.7cm}{\em 2.5~Summary of Available Software and Applications}........................................................& 7\\
\noindent {\scshape {3~Survey
    Outcomes}}.............................................................................................................& 8\\
\noindent {\scshape {4~Training of Future Computing and Software
    Experts}}.......................................& 9\\
\noindent {\scshape {5~Summary of Opportunities}}............................................................................................& 10\\
\\
\noindent {\scshape {1~Software Libraries and Tools Report:
    Introduction}}.........................................& 11\\
\noindent {\scshape {2~Prioritized
    Efforts}}........................................................................................................& 11\\
\hspace{0.7cm}{\em 2.1~Cross-Experiment
  Effort}..............................................................................................& 11\\
\hspace{0.7cm}{\em 2.2~Effort by
  Experiments}..................................................................................................& 12\\
\noindent {\scshape {3~Survey of Current Landscape}}....................................................................................& 13\\
\hspace{0.7cm}{\em 3.1~Forces Counter to Cross-Experiment
  Software}............................................................& 13\\
\hspace{0.7cm}{\em 3.2~Best Practices for
  Experiments}....................................................................................& 15\\
\hspace{0.7cm}{\em 3.3~Areas of
  Opportunity}...................................................................................................&
16\\
\noindent {\scshape {4~Event Processing Software Frameworks}}...............................................................& 16\\
\hspace{0.7cm}{\em 4.1~Description}..................................................................................................................&
16\\
\hspace{0.7cm}{\em 4.2~Gaudi}..........................................................................................................................&
17\\
\hspace{0.7cm}{\em 4.3~CMSSW and {\em art}}..........................................................................................................&
18\\
\hspace{0.7cm}{\em 4.4~IceTray}........................................................................................................................&
20\\
\hspace{0.7cm}{\em 4.5~Opportunities for Improvement}...................................................................................&
20\\
\noindent {\scshape {5~Software Development}}................................................................................................& 21\\
\hspace{0.7cm}{\em 5.1~Description}.................................................................................................................&
21\\
\hspace{0.7cm}{\em 5.2~Follow Free Software}...................................................................................................&
22\\
\hspace{0.7cm}{\em 5.3~Category Integration}...................................................................................................&
23\\
\hspace{0.7cm}{\em 5.4~Distributed Software Tools}..........................................................................................&
24\\
\hspace{0.7cm}{\em 5.5~Automate}....................................................................................................................&
25\\
\hspace{0.7cm}{\em 5.6~Opportunities for Improvement}...................................................................................&
25\\
\noindent {\scshape {6~Data Management}}..........................................................................................................& 26\\
\hspace{0.7cm}{\em 6.1~Definition}....................................................................................................................&
26\\
\hspace{0.7cm}{\em 6.2~Moving Data}...............................................................................................................&
26\\
\hspace{0.7cm}{\em 6.3~Metadata, Data Catalogs, and Levels of Data Aggregation}.........................................&
28\\
\hspace{0.7cm}{\em 6.4~Small and Medium Scale Experiments}........................................................................&
30\\
\hspace{0.7cm}{\em 6.5~Opportunities for Improvements}.................................................................................&
31\\
\end{tabular}

\begin{tabular}{lr}
\noindent {\scshape {7~Workflow and Workload Management}}...................................................................& 31\\
\hspace{0.7cm}{\em 7.1~The Challenge of the Three Domains}..........................................................................&
31\\
\hspace{0.7cm}{\em 7.2~Description}..................................................................................................................&
32\\
\hspace{0.7cm}{\em 7.3~Examples}.....................................................................................................................&
36\\
\hspace{0.7cm}{\em 7.4~Common Features}........................................................................................................&
37\\
\hspace{0.7cm}{\em 7.5~Intelligent Networks and Network Intelligence}.............................................................&
39\\
\hspace{0.7cm}{\em 7.6~Opportunities for Improvement}....................................................................................&
39\\
\noindent {\scshape {8~Geometry Information Management}}.........................................................................& 41\\
\hspace{0.7cm}{\em 8.1~Description}...................................................................................................................&
41\\
\hspace{0.7cm}{\em 8.2~Unified System}.............................................................................................................&
42\\
\hspace{0.7cm}{\em 8.3~Problems with CAD}......................................................................................................&
42\\
\hspace{0.7cm}{\em 8.4~Opportunities for Improvement}....................................................................................&
43\\
\noindent {\scshape {9~Conditions Databases}}.....................................................................................................& 43\\
\hspace{0.7cm}{\em 9.1~Description}...................................................................................................................&
43\\
\hspace{0.7cm}{\em 9.2~Basic Concepts}.............................................................................................................&
44\\
\hspace{0.7cm}{\em 9.3~Examples}......................................................................................................................&
45\\
\hspace{0.7cm}{\em 9.4~Opportunities for Improvement}....................................................................................&
45\\
\\
\noindent {\scshape {1~Systems Report:
    Introduction}}...................................................................................& 47\\
\noindent {\scshape {2~Computing Across HEP}}.................................................................................................& 47\\
\hspace{0.7cm}{\em 2.1~HEP Experimental
  Workflows}.....................................................................................& 47\\
\hspace{0.7cm}{\em 2.2~Computing for HEP Theory}........................................................................................& 52\\
\noindent {\scshape {3~Software Development: Incompatibility with the
    Systems Roadmap}}............& 54\\
\hspace{0.7cm}{\em 3.1~Cosmic Frontier: DES}................................................................................................& 54\\
\hspace{0.7cm}{\em 3.2~Energy Frontier: LHC Experiments}............................................................................& 55\\
\hspace{0.7cm}{\em 3.3~Intensity Frontier}........................................................................................................& 56\\
\hspace{0.7cm}{\em 3.4~Counter-Examples}.......................................................................................................& 56\\
\noindent {\scshape {4~Effects of Changing Technologies}}..........................................................................& 56\\
\hspace{0.7cm}{\em 4.1~Processors}...................................................................................................................& 56\\
\hspace{0.7cm}{\em 4.2~Software Challenges: Programmability versus Efficiency}............................................& 57\\
\hspace{0.7cm}{\em 4.3~Storage Hardware}........................................................................................................& 57\\
\hspace{0.7cm}{\em 4.4~Virtualization}..............................................................................................................& 58\\
\hspace{0.7cm}{\em
  4.5~Networking}..................................................................................................................&
59\\
\hspace{0.7cm}{\em 4.6~Non von Neumann Architectures}................................................................................& 60\\
\noindent {\scshape {5~The HEP Distributed Computing Environment}}.....................................................& 61\\
\hspace{0.7cm}{\em 5.1~Resources and Resource
  Provisioning}.........................................................................& 61\\
\hspace{0.7cm}{\em 5.2~HEP Applications and
  Networking}.............................................................................& 61\\
\hspace{0.7cm}{\em 5.3~Global Data Access}.....................................................................................................& 62\\
\hspace{0.7cm}{\em 5.4~Systems Data Analytics}..............................................................................................& 63\\
\hspace{0.7cm}{\em 5.5~Federated Identity Management}..................................................................................&
63\\
\\
\noindent {\scshape {References}}..........................................................................................................................& 64\\ 
\noindent {\scshape{Acronym
    Index}}...................................................................................................................&
68\\ 
\noindent {\scshape{Acknowledgments}}.............................................................................................................& 71\\ 
\noindent {\scshape{Disclaimer}}...........................................................................................................................& 72\\ 

\end{tabular}

\newpage


\setcounter{page}{1}

\begin{center}
{\bf\scshape{HEP-FCE Working Group Reports Summary}}
\end{center}

High energy physics relies critically on scientific computing,
simulations, and advanced data handling and analysis techniques for
scientific success across its broad program. As a result, major
funding for computing is provided to all sectors of HEP via
``vertical'' paths through each research or technology project.
Traditionally, HEP computing innovations and advances have been
developed within individual experiments or projects within the
confines of the current vertically funded model. While such advances
add tremendous value to the host experiment or project, these
contributions risk being lost to the rest of the HEP community, along
with any potential transfers to technology, without a horizontal
channel for easing exchanges and fostering innovation. Additionally,
as experimental data rates and volumes increase rapidly in an era of
constrained budgets, it is becoming increasingly apparent that a more
coherent response to these technological pressures is needed,
establishing the importance of adding a well-chosen ``horizontal''
component within the computing resources accessible to the field. The
P5 recommendation~\cite{p5}, echoes this reality: {\em Strengthen the
  global cooperation among laboratories and universities to address
  computing and scientific software needs, and provide efficient
  training in next-generation hardware and data-science software
  relevant to particle physics. Investigate models for the development
  and maintenance of major software within and across research areas,
  including long-term data and software preservation.}

The High Energy Physics Forum for Computational Excellence (HEP-FCE)
was established in the spring of 2014. It is DOE HEP's official
response to the P5 recommendation~\cite{p5} to strengthen global
cooperation among laboratories and universities and to address
significant scientific software and computing needs that have been
identified in a number of reports, including the Snowmass Community
Summer Study in 2013~\cite{snowmass}. The concept of the Forum was
originally proposed in a report from the Topical Panel Meeting on
Computing and Simulations in High Energy Physics convened in December
2013~\cite{hepcomp}.

Among the first tasks undertaken by the HEP-FCE was to establish three
working groups to focus on the current and future HEP computational
needs at the most fundamental level, with the aim of writing a report
detailing the current state of HEP computing and including key
observations for identifying opportunities that need to be exploited
in the future. These three working groups were
\begin{itemize}
\item Applications Software (leads -- Tom LeCompte and Zach Marshall) 
\item Software Libraries and Tools (leads -- Anders Borgland and Brett Viren)
\item Systems (hardware and systems software; lead -- Peter Nugent)
\end{itemize}
The scale of computing and data-intensive activities within HEP is
vast and covers efforts that include thousands of researchers at
hundreds of institutions worldwide, to collaborations of hundreds of
scientists, and finally, small groups and individual
investigators. Computational activities are similarly wide-ranging,
from science with datasets in the 100~Pbyte range and numerical
computations on the largest petascale systems available, to software
development and analysis on laptops. For these reasons, the technical
groups must necessarily address a large number of sub-topics.

The technical groups have interacted with the research community
across all of HEP and built on previous reports, the Snowmass White
Papers, and the work of the DOE HEP Topical Panel on Computing. While
the combined report identifies needs and opportunities for the field
as a whole, a subset of these are directly relevant in their impact on
the initial set of HEP-FCE activities.

The Applications Software Working Group was charged with surveying and
evaluating community software packages, large scale numerical
simulation codes, software being developed for next-generation
architectures, including HPC ports of experiment-specific codes, and
software management and distribution. The Working Group considered a
number of packages and toolkits that are in general use such as
Geant4~\cite{g4} for modeling the interactions of particles with
matter, the data analysis framework ROOT~\cite{root}, a wide variety
of event generators (general ones such as Pythia~\cite{pythia} and
Sherpa~\cite{sherpa}, and more specialized, such as
CORSIKA~\cite{corsika} for simulating cosmic rays). It was noted that
a wide variety of accelerator modeling codes exist (over 70) and are
used in different contexts. The evolution of the lattice QCD software
environment was also discussed. Given the diversity of the software
base, it was nevertheless felt that there were few obvious candidates
for a top-down driven consolidation -- for the most part new programs
were written to cover perceived inadequacies of older programs, so
many programs contain some unique functionality.  Bottom-up
consolidation where possible would be desirable, particularly
collaboration between existing groups to merge developments.

As part of the work of the Applications Software Working Group, a
survey was taken of members of the HEP community to ensure a broad
perspective in the report.  The respondents covered a range of
experiments and projects, though not all areas of the community were
represented.  Several dozen applications were discussed in the
responses. The response helped to clarify what was working well -- it
was felt in all cases that developers of the community software that
was in use were responsive and helpful -- and what was not, e.g.,
proper prioritization of software development when a small number of
developers must interact with a large community.

Because of the complex nature of many HEP computational activities,
there is widespread use of software libraries and tools. The Libraries
and Tools Working Group considered a number of topics ranging from
code management utilities, build/release/scripting/testing tools,
general purpose libraries, graphics packages, data management and
transfer tools, workflow and workflow management, and documentation
tools. The Working Group provided detailed information on a number of
selected areas (while identifying desirable properties) such as event
processing software frameworks, software development tools, and data
management tools. The Working Group concluded that sharing libraries
and tools across experiments would likely improve productivity. The
problem of the tendency to employ ``local'' solutions that actively
impede code resuse, maintenance, and flexibility was identified as a
major issue. Several other difficulties to be overcome were
discussed. These included lack of expertise, short-term commitments to
projects, parochial viewpoints, lack of openness, and errors in
setting the initial design directions. A set of best practices for
experiments were identified and discussed.

Finally, given the scale of modern software development, it was
considered beneficial to recognize a significant community-level
software commitment as a technical undertaking at times on par with
major detector R\&D. It was felt that recognizing people undertaking
major software developments would be extremely beneficial both to the
software projects and to the individual researchers, as recognition
through grant awards and similar processes is one metric by which
young scientists are judged. The important role of schools and
training programs in increasing awareness of modern software practices
was emphasized.

The Systems Working Group report considered the impact of changing
technologies on HEP computational practice. The topics covered include
processor technology, data access bottlenecks, virtualization,
resource provisioning, HEP applications and networking, optimization
of global data access, systems data analytics, and user
authentication.

After considering the large number of challenges and opportunities,
the Systems group identified two major issues for HEP computing; not
surprisingly these are 1) data storage and data access technologies,
and 2) efficient execution on future computer architectures. Issues of
data access and data storage are strongly intertwined. In the absence
of sufficiently fast networking one must increase the number of
data/compute hubs to maintain throughput. This increases the storage
that must be bought and the fine-grained nature of the computing
reduces the overall efficiency. Conversely, improvements in network
performance reduce the overall storage cache requirement and allow for
a smaller number of more powerful computing hubs, with an increase in
efficiency. As the data volume and the rate of data acquisition
increases, there is a corresponding increase of demand for
computational resources. Due to the constraints imposed by power
requirements, and the end of Dennard scaling~\cite{dennard}, computer
architectures are evolving in different directions~\cite{kogge}, none
of which are well-aligned with the current HEP software base, which
easily exceeds tens of millions of lines of code. Therefore, there is
an urgent need to develop new ideas and solutions that can transition
this software into a future state that is consistent with the hardware
evolution roadmap, a task that falls within the purview of the
software-related working groups. The future computing environment is
expected to be significantly dynamic -- at the level of the
computational resources, as well as from the point of view of resource
provisioning and resource utilization -- and will place concomitant
demands on networking. The resulting opportunities and challenges were
considered in the report.

The HEP-FCE Working Group reports considered the situation across all
of HEP and their conclusions apply to the field as a whole, including
modes of software development, best practices for HEP experiments,
hardware issues, and other, large-scale issues. Based on the findings
and conclusions in the three reports, a number of (limited-scope)
initial HEP-FCE activites have been identified. These include work in:
\begin{itemize}
\item Next-generation architectures and HPC/supercomputer applications
for HEP experiments
\item Data-intensive/cloud computing (virtualization/containers)
\item  Cross-cut software development
\item High-speed networking (large-scale data transfers in production mode)
\item ASCR/HEP interactions and workshops
\item HEP-FCE infrastructure support and community development

\end{itemize}

\newpage

\begin{center}
{\bf\scshape{1~~~Applications Software Report: Introduction}}
\end{center}

The Applications Software Working Group was charged with surveying and
evaluating community software packages, large scale numerical
simulation codes, software being developed for next-generation
architectures, including HPC ports of experiment-specific codes, and
software management and distribution.  This charge was in part a
response to the P5 panel's recommendation 29~\cite{p5}, to
``strengthen the global cooperation among laboratories and
universities to address computing and scientific software needs, and
provide efficient training in next-generation hardware and
data-science software relevant to particle physics. Investigate models
for the development and maintenance of major software within and
across research areas, including long-term data and software
preservation.''

This document first lays out a number of the packages in common use.
In general, we do not find any obvious place or need for a software
down-select; the community has done well even without deliberate or
conscious organization to manage its resources.  The results of a
survey of the community are discussed, and several suggestions are
made towards improving training of young scientists in software
development and helping to ensure that reasonable career paths are
available to physicists who are computing-minded.

\medskip

\begin{center}
{\bf\scshape{2~~~Packages in Common Use}}
\end{center}

There are a number of programs, packages and toolkits in broad use.
These are under active development, and in most cases there are
multiple possible directions for development.  A partial list follows.

\medskip

\noindent{\bf\scshape{2.1~Experiment and Phenomenology}}\\
Geant4 is a toolkit for modeling the interaction of particles with
matter.  It is used in varying degrees by virtually all experiments.
It is exceptionally flexible, and most experiments assign physicists
to adapting this flexibility to their specific cases.  For example,
for SuperCDMS~\cite{supercdms} two Geant4 developers are also members
of the collaboration.  The Geant4 toolkit includes most features
necessary for SuperCDMS backgrounds simulations, with the exception of
simulating the phonons and solid-state charge carriers which
constitute their signal.  However, the toolkit is sufficiently general
that, with the assistance of Geant4 developers, they have been able to
incorporate those features themselves.  Development efforts range from
multithreading, use of coprocessors, vectorization, and general
performance improvements, to improved physics modeling. The extension
of the physics models to the high energies required by the 14 TeV LHC
and the future colliders, as well as to low energies for
low-background experiments and other applications, are also ongoing.
Other specialized codes, including FLUKA~\cite{fluka},
MCNPX~\cite{mcnpx}, and SHIELD-HIT~\cite{shieldhit} tend to be used
for low-energy applications or background modeling for low-background
experiments.

ROOT is a data analysis framework used for individual analyses, but
built on a set of libraries that are used more generally within the
experiments (along with new aspects like PROOF and
XRootD~\cite{xrootd}, and derivative works like RooFit, RooStats,
PyROOT and ROOTPy).  It has recently had a major release, ROOT6, that
replaces the previous C++ interpreter Cint with Cling.  The
experiments are assessing this update and evaluating the best time to
make the change.

A wide variety of event generators exist, ranging from the general
(Pythia, Sherpa) to very specialized (CORSIKA, which simulates cosmic
rays).  Periodically there is talk of consolidation, but the same
conclusions are drawn: the specialized ones are unique, and the
general ones take very different approaches.  This difference makes
them both valuable in understanding the effect of these different
approaches and also difficult to merge.  There has been a slow move
towards automatic frameworks for higher-order event generation
(e.g. Sherpa 2 and aMC$@$NLO) in order to replace some of the
specialized generators that provided state-of-the-art accuracy when
they were written.  This trend will naturally continue, with new
generators coming in at high accuracy (e.g., Top++ at NNLO) and
specialized lower-order generators falling out of use as automatic
codes surpass them in speed, configurability, and ease of use.

For phenomenologists, very fast, publicly available simulation
software like the Pretty Good Simulation (PGS)~\cite{pgs} and
Delphes~\cite{delphes} are indispensable for testing new models of
physics beyond the standard model by recreating cut flows from
experimental searches, just as frameworks like Rivet~\cite{rivet} and
Professor~\cite{professor} are very helpful for testing new models
against measurements and for the tuning of event generators.  These
programs are often privately patched to improve the accuracy of some
difficult-to-model effects like b-quark tagging.  It would be helpful
for the theory community in general if these private patches were made
publicly available.  Some sort of community development would have to
be undertaken in order to make this possible, as happened to some
degree during the Snowmass exercises recently.  These developments
would serve as an excellent test case for how to centrally improve
these programs by incorporating user modifications.

{\em art} is a framework for new and (by the scale of collider
experiments) small experiments at the Intensity Frontier.  This is
presently a Fermilab product and used for Fermilab experiments, but it
has the potential to grow in scope.  (As a historical footnote, ATLAS
and LHCb initially used the same framework, although they diverged
over time to meet each of the experiments' needs.) It is discussed in the
context of the Tools and Libraries groups.

\medskip

\noindent{\bf\scshape{2.2~Accelerator Physics}}\\
Many computer simulation codes have been developed (over 70 worldwide)
for the modeling of particle accelerators and beam transport. There
has been little coordination of the development of the accelerator
physics codes whose aggregate involves a mix of complementarity and
duplication, and they are not all actively developed and maintained.

Many of the codes have been developed by a single developer (often a
physicist) for a specialized purpose or accelerator. Several
multi-physics frameworks were developed by small teams, some in large
part with the support of DOE's Scientific Discovery through Advanced
Computing (SciDAC) program~\cite{scidac}, and are capable of
incorporating many physics models. A substantial fraction of the codes
is serial, but a number of the codes have been ported to parallel
computers and some are capable of handling massive parallelism. A
small fraction of the codes were ported to GPUs. Many of the codes are
written in FORTRAN, C or C++, with a growing number combining the
compiled language modules (for number crunching) with a Python
scripting interface.

A list of major U.S. accelerator codes, frameworks and toolkits (may
not be exhaustive, and commercial codes are mostly not included):
\begin{itemize}
\item ACE3P (SLAC): Omega3P, Pic3P, S3P, T3P, Track3P, TEMP3P
\item BLAST (LBNL): BeamBeam3D, Impact, MaryLie/IMPACT (also U. Md,
  Tech-X), Posinst, Warp (also LLNL/U. Maryland) 
\item LAACG (LANL): Parmela, Parmila, Poisson/Superfish, Parmteq,
  Trace
\item BMAD (Cornell U.)
\item MARS, Synergia (FNAL)
\item COSY (MSU)
\item G4Beamline (Muons Inc.)
\item Elegant, Track (Argonne)
\item Orbit/PyOrbit (ORNL)
\item Osiris, QuickPIC (UCLA)
\end{itemize}

Until now, the development of accelerator codes has been left to
projects without the mandate and programmatic funding for
coordination, distribution and user support. While this is adequate
for the development of relatively small-scale codes on targeted
applications, a more coordinated approach is needed to 1) enable well
supported multi-physics codes with user bases that extend beyond
individual projects, 2) leverage crosscutting activities (e.g.,
porting codes to many-core or GPU architectures). It is however
desirable to capitalize on the existing pool of codes, which represent
a significant investment from the community, and to avoid disruption
to the users and developers by adopting an incremental (near
adiabatic) approach for transitioning from the existing collection of
codes to a modular ecosystem of interoperable components that
facilitate cooperation and reuse. It is also important that
innovations in algorithms, which is a strength of this community, is
not hindered by the transition. Such an approach is being initiated by
the new Consortium of Advanced Modeling of Particle Accelerators
(CAMPA~\cite{campa}).

\medskip

\noindent{\bf\scshape{2.3~Lattice QCD}}\\
There is a long standing software effort within the US lattice field
theory community funded through the SciDAC program, and there are a
number of community codes that are freely available.  Traditionally,
the emphasis has been on QCD, i.e., SU(3) gauge theory with various
formulations for the quarks.  More recently, there has been
significant work on theories other than QCD, which may be relevant for
beyond the standard model physics, including supersymmetric theories.
High performance has always been a necessity and a point of pride for
this community.  For many years there was a fairly stable programming
environment with message passing between nodes and single core
performance optimized via libraries.  The programming environment has
become increasingly challenging as we deal with new architectures such
as GPUs and many-core chips.  These require two or three levels of
parallelism and very careful consideration of the data layout.

The USQCD Collaboration, which is responsible for the SciDAC software
effort, has developed three levels of libraries to support the
application codes~\cite{usqcd}.

Level 1 consists of a basic linear algebra library, QLA, a message
passing library, QMP, and a multi-threading library, QMT. At level 2
are the C and C++ libraries, QDP and QDP++, which contain data
parallel operations that combine linear algebra operations with shifts
of data between grid points; the LIME library, C-LIME, specialized for
QCD; and Bagel QDP, which uses Peter Boyle's Bagel package for
optimized linear algebra. Level 3 contains highly optimized packages
for solvers of various types that have a standard interface so they
can be called by the various community application codes.  Among the
libraries is one designed for NVIDIA GPUs called QUDA.

The original application libraries for QCD are Chroma (Jefferson Lab),
CPS (Columbia, BNL, UKQCD), and MILC (MILC Collaboration).  Recently,
the high level libraries FUEL (Argonne) and QLUA (MIT) have been
developed to enable applications with a wider variety of gauge groups
and fermion representations.  Continuing development will be needed to
exploit current and future hardware.  It is crucial not to lose
expertise in GPU computing and to continue to develop expertise in
many-core computing.  Furthermore, algorithm development cannot be neglected as it
has the potential to yield much greater benefits than code
optimization.

\medskip

\noindent{\bf\scshape{2.4~Computational Cosmology}}\\
Cosmological simulations can be classified into two types: 1)
gravity-only N-body simulations, and 2) hydrodynamic simulations
that also incorporate gasdynamics, sub-grid modeling, and feedback
effects. Because gravity dominates on large scales, and dark matter
outweighs baryons by roughly a factor of five, N-body simulations
provide the bedrock on which all other techniques rest. Parallel
numerical implementations can be purely particle-based or
particle/grid hybrids. Several post-processing strategies exist to
incorporate additional physics on top of the basic N-body
simulation. The key shortcoming is that much of the physics of the
baryonic sector cannot be treated directly. Whenever the dynamics of
baryons is important, gasdynamic, thermal, and radiative processes --
among others -- must be incorporated along with sub-grid modeling of
processes such as star formation and local feedback mechanisms. Such
simulations are substantially more complex and difficult to carry out,
and at present they are limited to volumes significantly smaller than
full survey volumes. `Gastrophysics' is added to N-body simulations
via either grid-based adaptive mesh refinement (AMR) solvers or via
particle-based methods such as smoothed-particle hydrodynamics (SPH).
With partial support from the SciDAC program, three large-scale
cosmology codes are being used -- the extreme-scale N-body code
HACC~\cite{hacc} and the state of the art cosmological hydrodynamics
codes, ART and Nyx~\cite{Nyx}.

HACC is targeted at exploiting next-generation architectures and can
run on CPU, CPU/GPU, and many-core systems. ART and Nyx will need
refactoring to run on many-core systems (there is currently no plan to
use GPUs). As in the case of lattice QCD, developing and maintaining
computational and algorithmic expertise in this area is an essential
requirement. 

\medskip

\noindent{\bf\scshape{2.5~Summary of Available Software and
    Applications}}\\
As shown, the number of programs is rather large.  There is no obvious
candidate for a top-down driven consolidation: new programs were
written to cover perceived inadequacies of older programs, so many or
all programs contain some unique functionality.  We encourage
bottom-up consolidation where possible, particularly collaboration
between existing groups to merge developments.  Ensuring that some
conferences like CHEP provide an opportunity for the announcement and
discussion of new software package for HEP applications is critical
both for experiments to have a good view of the available options and
for developers to gain recognition for their work.

Over the last several years, quite a few ``wrapper'' packages (e.g.,
rootpy~\cite{rootpy}) or HistFitter~\cite{histfitter}) have been
developed around some of the larger software applications that are in
wide use (in these cases, ROOT/PyROOT and RooFit/RooStats).  It would
be very useful for the larger collaborations supporting these software
applications to carefully examine these wrapper packages to see what
functionality could be integrated back into the main package, and what
led people to develop the wrappers in the first place.  In some cases
the differences may simply be a matter of taste; in other cases it
could be that the developments are useful to a much wider community,
and simply because they are maintained separately the wrapper packages
are generally not as well known.

\medskip

\begin{center}
{\bf\scshape{3~~~Survey Outcomes}}
\end{center}

A survey was taken of members of the HEP community to ensure a broad
perspective in this report.  The respondents covered a range of
experiments and projects, though not all areas of the community were
represented.  Several dozen applications were discussed in the
responses.

The following questions were asked in the community survey:
\begin{itemize}
\item What software packages does your collaboration use that were not
  internally developed, or that were internally developed but are now
  used extensively beyond the collaboration? 
\item Are the packages you listed still actively maintained?
\item Are the software package developers responsive?
\item Are new developments, beyond just bugfixes, going into these
  software packages? 
\item Do you find that the software in the package has sufficient
  functionality for your use case? Furthermore, if there are new
  developments, do you anticipate these meeting your needs over the
  next few years? 
\item Do you find that the software package has significant
  performance penalties that you need to work around, or which are
  causing bottlenecks in your own code performance? 
\end{itemize}

In every case, it was reported that the developers of the community
software that was in use were responsive and helpful.  Some packages
do not appear to be actively maintained, but in those cases the
functionality has been taken up by another project, so these efforts
do not need to be restarted.  Development is steady in the major
packages, but major development may be disruptional -- as one
responder put it, ``We're so used to it that big changes might be
counterproductive.''  Another pointed out that ``to get the physics
updates to a package, we often have to accept breaking interface
changes, etc.''  Development is still ``driven by collaboration
needs,'' according to another respondent, but this may be a part of
the problem: when a small number of developers are fielding requests
from a large user-base, it can be difficult to weigh the importance of
one group of requests against another without more regular and public
interactions.  As one respondent put it, the development happens best
``when we know what to ask.''  With regards to a number of these
concerns, the Geant4 collaboration has an excellent model:
requirements, requests, and bug reports are tracked in a public
manner, and there are regular (several times per year) meetings at
which users are able to raise concerns and at which developers report
new features to inform the larger community.  This also serves to
guide the development process, as users have a clear opportunity to
respond early to major developments (whether positively or negatively)
and to learn what issues others have seen and overcome.

A few opportunities for common software projects were identified.  One
of these is a common geometry project.  At the moment, ROOT, Geant4,
DD4HEP, USolids, and GDML all provide very similar functionality in
terms of geometry description, and other experiments (e.g. ATLAS's
GeoModel) or software packages (e.g. FLUKA) provide yet another
geometry description style and language.  These different codes have
various performance features, and in some cases several have been
integrated inside of a single package (e.g., Geant4 is able to run
with several different geometry engines).  It would be beneficial to
see how much of these different packages could be integrated within a
smaller number of efforts.

In most cases, survey respondents reported that software or library
performance was adequate for their needs, or that they were able to
work around any major performance bottlenecks.  In all cases where the
software had performance problems, the alternative solutions on the
market did not provide adequate functionality or had their own set of
disadvantages.  Several codes and groups were complimented for their
consistent performance improvements and for their use of modern
hardware architectures.

\medskip

\begin{center}
{\bf\scshape{4~~~Training of Future Computing and Software Experts}} 
\end{center}

In recent years, some software projects have become just as complex,
and come to require comparable person-hours and resources, as major
hardware research and development projects.  As yet, however, the
community has not fully recognized the analogous need with analogous
funding opportunities or future job prospects.  Traditional academic
environments have recognized the importance of technical hardware work
in combination with strong physics programs. Given the scale of modern
software development, it would be beneficial to recognize a
significant community-level software commitment as a technical
undertaking at times on par with major detector R\&D.  Of course,
there are some differences between the two, in terms of the physical
nature of hardware (e.g., it is easier to show a machine that has been
purchased or a new chip that has been developed) and the casual nature
of some software (e.g. physics analysis code is not the appropriate
level of effort).  But we believe that recognizing people undertaking
major software developments will be hugely beneficial both to the
software projects and to the researchers involved, as recognition
through grant awards and similar processes is one metric by which
young scientists are judged.  Such opportunities, provided through the
HEP-FCE or by the DOE Office Of Science directly, would help improve
appreciation for expertise that is often unrecognized in traditional
academic environments.  The HEP-FCE has already put forward the notion
of advertising the work of early-career people in the field on the
web, which is an excellent start.

In a similar vein, summer schools and other programs for the training
of graduate students and young scientists have traditionally included
some hardware training aspect.  If a software expert's long-term job
prospects are comparable to scientists with an equivalent hardware
expertise, then the community will benefit greatly from additional
training in software issues.

These schools and training programs should be augmented to ensure
that, just as students rarely reach graduation without some exposure
to hardware techniques and issues, students in the field are afforded
comparable training in modern software development techniques,
including programming on new advanced hardware platforms.  The HEP-FCE
is an obvious contact point for finding experts to get involved in the
development of such programs.  These should be augmented by special
workshops and training sessions, organized in part by the HEP-FCE, for
the training of future experts.  Here OpenLab~\cite{openlab} is an
example of an organization with significant commitment to hosting
training and tutorials in important modern software topics.  Topics
for such sessions could include:
\begin{itemize}
\item Co-processor and GPU use, and the porting of widely-used
  software packages 
\item Low-level performance profiling and common beneficial design
  patterns 
\item Novel pattern-recognition algorithms
\end{itemize}

\medskip

\begin{center}
{\bf\scshape{5~~~Summary of Opportunities}} 
\end{center}

We do not find significant opportunities for software package
down-selects at this moment in time.  The software and applications in
use in the HEP community have been evolving and will continue to as
new possibilities are opened by new technologies and advances in
theoretical understanding.  There are several large software packages
in wide use, including ROOT and Geant4.  These have generally done
well responding to developers needs, but more coordination and
response might be constructive in some cases.

There are significant opportunities for better training and
advancement for scientists within the software development community.
Modern software development should be recognized as having comparable
complexity to many hardware R\&D projects, and the agencies could help
to ensure this recognition.  Improving student training in good
programming practices on modern platforms would also help the field in
general, and there are several opportunities in this area.

\newpage

\begin{center}
{\bf\scshape{1~~~Software Libraries and Tools Report: Introduction}}
\end{center}

This report summarizes the deliberations of the HEP-FCE Software
Libraries and Tools Working Group. The charge to the Working Group
includes topics such as:
\begin{itemize}
\item Code management utilities
\item Build/release/scripting/testing tools
\item Documentation tools
\item Graphics packages
\item General purpose libraries (I/O, statistical analysis, linear algebra)
\item Data management and transfer tools
\item Workflow and Workload management
\end{itemize}

The focus of the this Working Group report is on software libraries
and tools, however, the breadth and depth of work relevant to HEP in
this area is far too extensive to provide complete coverage in this
document. Instead of attempting to be comprehensive, the Working Group
has considered only a sampling, hopefully representative, of the
possible projects and areas. Omissions are not intended to be
interpreted as positive or negative reflections on those projects or
areas.  In the following sections we give a prioritized list of
technical activities with suggested scoping and deliverables that can
be expected to provide cross-experiment benefits. The remaining bulk
of the report gives a technical survey of some specific ``areas of
opportunity'' for cross-experiment benefit in the realm of software
libraries and tools. This survey serves as support for the prioritized
list. For each area we describe the ways that cross-experiment benefit
is achieved today, as well as describe known failings or pitfalls
where such benefit has failed to be achieved and which should be
avoided in the future. For both cases, we try to give concrete
examples. Each description then ends with an examination of what
opportunities exist for improvements in that particular area.

\medskip

\begin{center}
{\bf\scshape{2~~~Prioritized Efforts}}
\end{center}

\noindent{\bf\scshape{2.1~Cross-Experiment Effort}}\\   
\begin{enumerate}
\item Various detailed ``opportunities'' listed in the following
  survey sections call out the need for further work to be carried out
  in some detail by technical working groups. These are needed to
  better understand the nature of a specific problem shared across
  many experiments, formulate requirements for -- and in some cases --
  design and implement solutions. Such working groups should be
  organized using suitable expertise from the HEP software community.
\item Packages or frameworks (or significant subsets) which have
  proven popular (used by more than one experiment) and useful should
  be considered for cross-experiment support, especially in terms of
  providing support for easy adoptability (setup and install by other
  experiments, on other O/S platforms) and documentation (detailed
  guides and non-experiment-specific manuals).
\end{enumerate}

\noindent{\bf\scshape{2.2~Effort by Experiments}}\\   
Throughout the following survey sections, a number of best practices
and pitfalls relevant to the development and use of software libraries
and tools by individual experiments were identified. First, some
generalities were identified:

\begin{enumerate}
\item New experiments should not underestimate the importance of
  software to their success. It should be treated as a major subsystem
  at least on par with other important aspects such as detector
  design, DAQ/electronics, civil construction, etc.
\item Experiments should understand the pitfalls listed in
  Section~3.1. New experiments should plan and implement mechanisms to
  avoid them and existing experiments should reflect on which ones may
  apply and develop ways to address them. Likewise, the best practices
  listed in Section~3.2) should be considered. New experiments should
  attempt to follow them and if practical and beneficial, existing
  experiments should seek to make the changes needed to implement
  them.
\end{enumerate}

The remaining Sections below contain surveys of select areas of
software libraries and tools in HEP. For each we list a summary of
aspects that make for success in the associated area.

\begin{itemize}
\item Aspects of successful Event Processing Software Frameworks
  include: those with flexible (possibly hierarchical) notions of
  ``events'', those that are easily adoptable by new experiments, are
  well-documented, have dynamically configurable (possibly scriptable)
  configuration parameter sets and are modular and efficient (e.g.,
  allow C++ like modules for low-level operations combined with a
  scripting layer like Python for flexible higher level control).
\item Aspects of successful Software Development tools include: those
  that follow licence-free availablity and free-software distribution
  models, those that include code repositories, build systems that
  work on a variety of platforms with a small number of clearly
  defined base element dependencies (i.e., C compiler, compression
  library, specific version of Python) and those with release
  configuration systems with versioning that understand a variety of
  platforms; those that support automatic continuous integration and
  regression testing; those that have open documentation updating and
  bug-reporting and tracking.
\item Aspects of successful Data Management tools include: those that
  are inherently modular and avoid tight couplings to either specific
  technologies or to other parts of the computing ecosystem, in
  particular to the Workload Management System and the Metadata
  Catalogs; those that, while being complex and originallly developed
  for large scale operations, at for example the LHC, may be
  simplified and downscaled for use by smaller experiments with
  minimal manpower and technical expertise.
\item Aspects of successful Workflow and Workload Management tools
  include: those that understand the distinction between and support
 flexible, efficient interaction between workflow, workload, and data
  management aspects of a system; those that make efficient use of
  resources (CPU, RAM-memory, disk, network, tape) for processing in
  parallel; those that allow granular, multi-level monitoring of
  status; those that handle error cases effectively and inclusively;
  those that are properly scaled to the size of the experiment.
\item Aspects of successful Geometry Information Management tools
  include: those that follow or set widely used standards for
  representation of geometric information; those that follow standards
  for visualization.
\item Aspects of successful Conditions Database tools include: those
  that allow standardized, experiment-wide access to representative or
  specific event conditions so that realistic simulations or
  statistics can be generated by users without detailed knowledge of
  detectors or specific event.
\end{itemize}

\medskip

\begin{center}
{\bf\scshape{3~~~Survey of Current Landscape}}
\end{center}

This Section presents a general overview of the current landscape of
HEP libraries and tools. First we list general patterns that run
counter to cross-experiment sharing. Secondly, we give a prioritized
list of beneficial activities.

\medskip

\noindent{\bf\scshape{3.1~Forces Counter to Cross-Experiment
    Software}}\\   
Sharing software libraries and tools between experiment more
frequently than is currently done is expected, by the group, to
increase overall productivity. Independent of cross-experiment
sharing, designing and implementing software in a more general manner
is expected to be beneficial. The Working Group identified some
reasons why such general use software is not as predominant as it
could be.

\medskip

\noindent{\bf\em{3.1.1} Up-front Effort}\\
Designing and implementing software to solve a general problem instead
of the specific instance faced by one experiment can take more effort
initially. Solving ``just'' the problem one immediately faces is
cheaper in the immediate time scale. If the problem is short-lived and
the software abandoned, this strategy can be a net benefit. What is
more often the case, fixes to new problems compound the problem and
the software becomes either brittle and narrowly focused, increasingly
diffcult to maintain, and ever less able to be extended.

\medskip

\noindent{\bf\em{3.1.2} Lack of Expertise}\\
Physicists have always been multidisciplinary, covering all aspects of
an experiment from hardware design, bolt turning, operations, project
management, data analysis and software development. As data rates have
increased, algorithms have become more complex, and networking,
storage and computation technology more advanced, the requirements for
a physicist to be a software developer have become more challenging to
meet, while maintaining needed capabilities in the other
disciplines. As a consequence, some experiments -- especially smaller
ones -- lack the software expertise and knowledge of what is available
needed to develop general software solutions, or adopt existing
ones. This leads to the same result of solving ``just'' the immediate
problem and associated consequences described above.

\medskip

\noindent{\bf\em{3.1.3} Ignoring Software Design Patterns}\\
A specific example of lack of expertise manifests in developers who
ignore basic, tried and true, software design patterns. This can be
seen in software that lacks any notion of interfaces or layering
between different functionality. Often new features are developed by
finding a spot that ``looks good'' and pasting in some more code to
achieve an immediate goal with no understanding of the long-term
consequences. Like the ``up-front'' costs problem, this strategy is
often rewarded as the individual produces desired results quickly and
the problem that this change causes does not become apparent until
later.

\medskip

\noindent{\bf\em{3.1.4} Limited Support}\\
Some experiments have a high degree of software expertise. These
efforts may even naturally produce software that can have some
cross-experiment benefit. However, they lack the necessary ability to
support their developers to make the final push needed to offer that
software more broadly. In many cases they also do not have the ability
to assure continued support of the software for its use by others. In
the best cases, some are able to provide support on a limited or best
effort basis. While this helps others adopt the software, it still
leaves room for improvements. A modest amount of expert time can save
a large amount of time of many novices.

\medskip

\noindent{\bf\em{3.1.5} Transitory Members}\\
Many software developers in an experiment are transitory. After
graduate students and post-docs make a contribution to the software
development and the experiment in general they typically move on to
other experiments in the advancement of their careers. In part, this
migration can help disseminate software between experiments but it
also poses the problem of retaining a nucleus of long-term knowledge
and support around the software they developed.

\medskip

\noindent{\bf\em{3.1.6} Parochial View}\\
In some cases, beneficial software sharing is hampered by experiments,
groups, labs, etc which suffer from the infamous ``not invented here''
syndrome. A parochial view leads to preferring solutions to come from
within the unit rather than venturing out and surveying a broader
landscape where better, more general solutions are likely to be
found. Parochialism compounds itself by making it ever more difficult
for motivated people to improve the entrenched systems by bringing in
more general solutions.

\medskip

\noindent{\bf\em{3.1.7} Discounting the Problem}\\
There is a tendency for some physicists to discount software and
computing solutions. The origin of this viewpoint may be due to the
individual having experience from a time where software and computing
solutions were indeed not as important as they are now. It may also
come as a consequence of that person enjoying the fruits of high
quality software and computing environments and being ignorant of the
effort needed to provide and maintain them. Whatever the origin,
underestimating the importance of developing quality software tools
leads to inefficiency and lack of progress.

\medskip

\noindent{\bf\em{3.1.8} Turf Wars}\\
Software development is a personal and social endeavor. It is natural
for someone who takes pride in that work to become personally attached
to the software they develop. In some cases this can cloud judgment
and lead to retaining software in its current state while it may be
more beneficial to refactor or discard and reimplement. What are
really prototypes can become too loved to be replaced.

\medskip

\noindent{\bf\em{3.1.9} Perceived Audience and Development Context}\\
The group made the observation that cues from the audience for the
software and the context in which it is developed lead to shifts in
thinking about software design. For example, the resulting designs
tend to be more narrowly applicable when one knows that the code will
be committed to a private repository accessible only by a single
collaboration. On the other hand, when one is pushing commits to a
repository that is fully accessible by a public audience one naturally
thinks about broader use cases and solutions to more general problems.

\medskip

\noindent{\bf\em{3.1.10} Disparate Communications}\\
Different experiments and experiment-independent software projects
have differing means of communicating. Technical support, knowledge
bases, software repositories, bug trackers, release announcements are
all areas that have no standard implementation. Some groups even have
multiple types of any of these means of communication. Independent of
this, different policies mean that not all information may be publicly
available. These all pose hurdles for the sharing of software between
groups.

\medskip

\noindent{\bf\em{3.1.11} Design vs. Promotion}\\
For general purpose software to be beneficial across multiple
experiments it needs at least two things. It needs to be well designed
and implemented in a way that is general purpose. It also needs to be
promoted in a way so that potential adopters learn of its
suitability. Often the set of individuals that excel at the former and
excel at the latter have little overlap.

\medskip

\noindent{\bf\em{3.1.12} Decision Making}\\
An experiment's software is no better than the best software expert
involved in the decision making process used to provide it. And it's
often worse. Decision making is a human action and as such it can
suffer from being driven by the loudest argument and not necessarily
the one most sound. Many times, choices are made in a vacuum lacking
suitable opposition. At times they are made without a decision-making
policy and procedures in place or ignored if one exists, or if
followed, without sufficient information to make an informed
decision. Politics and familiarity can trump rationality and quality.

\medskip

\noindent{\bf\em{3.1.13} Getting off on the Wrong Foot}\\
There is often no initial review of what software is available when a
new experiment begins. Frequently a physicist charged with software
duties on an experiment will jump in and begin to do things the way
that they were done in their last project, thus propagating and baking
in inefficiencies for another generation.  No time will be spent to see
what has changed since an earlier experiment's software design, and
whole evolutions in ways of thinking or recently available tools
updates may be missed.

\medskip

\noindent{\bf\scshape{3.2~Best Practices for Experiments}}    

\medskip

\noindent{\bf\em{3.2.1} Look Around}\\
New experiments should survey and understand the current state of the
art for software libraries and tools (and Systems and Applications
Software as covered by the other two working groups). Periodically,
established experiments should do likewise to understand what
improvements they may adopt from or contribute to the
community. Experts from other experiments should be brought in for
in-depth consultation even in (especially in) cases where the
collaboration feels there is sufficient in-house expertise.

\medskip

\noindent{\bf\em{3.2.2} Early Development}\\
There are certain decisions that if made early and implemented can
save significant effort in the future.  Experiments should take these
seriously and include them in the conceptual and technical design
reports that are typically required by funding agencies. These include
the following:

\medskip

\noindent{\bf data model}\\
Detailed design for data model schema covering the stages of data
production and processing including: the output of detector simulation
(including ``truth'' quantities), the output of ``raw'' data from
detector DAQ and the data produced by and used as intermediaries in
reconstruction codes.

\medskip

\noindent{\bf testing}\\
Unit and integration testing methods, patterns and granularity. These
should not depend on or otherwise tied to other large scale design
decisions such as potential event processing software frameworks.

\medskip

\noindent{\bf namespaces}\\
Design broad-enough namespace rules (unique
filenames, event numbering conventions, including re-processed event
version tags) to encompass the entire development, operations and
legacy aspects of the experiment, which may span decades in time and
have worldwide distributed data stores.  Filenames, or, in larger
experiments, the meta-system which supports file access and movement,
should have unique identifiers not just for given events or runs at a
single location, but even if a file is moved and mixed with similar
files remotely located (i.e., filename provenance should not rely upon
directory path for uniqueness). One should be able to distinguish
development versions of files from production versions. If the same
dataset is processed multiple times, the filenames or other metadata
or provenance indicators should be available that uniquely track the
processing version. The same goes for code: software versions must be
tracked clearly and comprehensively across the distributed experiment
environment (i.e., across multiple institutions, experiment phases and
local instances of repositories).

\medskip

\noindent{\bf scale}\\
Understand the scale of complexity of the software, its
development/developers. Determine if an event processing framework is
needed or if a toolkit library approach is sufficient or maybe if
ad-hoc development strategies are enough.

\medskip

\noindent{\bf metadata}\\
Determine what file metadata will be needed across the entire efforts
of the collaboration. Include raw data and the requirements for its
production as well as simulation and processed data. Consider what
file metadata will be needed to support large scale production
simulation and processing. Consider what may be needed to support
ad-hoc file productions by individuals or small groups in
collaboration.

\medskip

\noindent{\bf\scshape{3.3~Areas of Opportunity}}\\    
Each of the following sections focus on one particular {\em area of
  opportunity} to make improvements in how the community shares
libs/tools between experiments. In each area of opportunity we
present:

\begin{itemize}
\item A description of the area.
\item A number of case studies of existing or past software libraries
  and tools including concrete examples of what works and what does
  not.
\item Specific aspects that need improvement and an estimation of what
  efforts would be needed to obtain that.
\end{itemize}

\medskip

\begin{center}
{\bf\scshape{4~~~Event Processing Software Frameworks}}
\end{center}

\noindent{\bf\scshape{4.1~Description}}\\ 
A software framework abstracts common functionality expected in some
domain. It provides some generic implementation of a full system in an
abstract way that lets application-specific functionality to be added
through a modular implementation of framework interfaces.

Toolkit libraries provide functionality addressing some domain in a
form that requires the user-programmer to develop their own
applications. In contrast, frameworks provide the overall flow control
and main function requiring the user-programmer to add application
specific code in the form of modules.

In the context of HEP software, the terms ``event'' and ``module'' are
often overloaded and poorly defined.  In the context of software
frameworks, an ``event'' is a unit of data whose scope is dependent on
the ``module'' of code which is processing. In the context of a code
module that generates initial kinematics, an event is the information
about the interaction. In a module that simulates the passage of
particles through a detector, an event may contain all energy
depositions in active volumes. In a detector electronics simulation,
it may contain all signals collected from these active volumes. In a
trigger simulation module, it would be all readouts of these signals
above some threshold or other criteria. At this point, data from real
detectors gain symmetry with simulation. Going further, data
reduction, calibration, reconstruction and other analysis modules each
have a unique concept of the ``event'' they operate on. Depending on
the nature of the physics, the detector, and the follow-on analysis,
every module may not preserve the multiplicity of data. For example, a
single interaction may produce multiple triggers, or none.

With that description, an event processing software framework is
largely responsible for marshalling data through a series (in general
a directed and possibly cyclic graph) of such code modules which then
mutate the data. To support these modules the framework provides
access to external services such as data access, handle file I/O,
access to descriptions of the detectors, provide for visualization or
statistical summaries, and databases of conditions for applying
calibrations. The implementation of these services may be left up to
the experiment or some may be generically applicable. How a framework
marshalls and associates data together as an event is largely varied
across different HEP experiments and may be unique for a given data
collection methodology (beam gate, online trigger, raw timing, etc).

\medskip

\noindent{\bf\scshape{4.2~Gaudi}}\\    
The Gaudi event processing framework~\cite{gaudi1, gaudi2} provides a
comprehensive set of features and is extensible enough that it is
suitable for a wide variety of experiments. It was conceived by LHCb
and adopted by ATLAS and these two experiments still drive its
development. It has been adopted by a diverse set of experiments
including HARP, Fermi/GLAST, MINER$\nu$A, Daya Bay and others. The
experience of Daya Bay is illuminating for both Gaudi specifically and
for more general issues of this report.

First, the adoption of Gaudi by the Daya Bay collaboration was greatly
helped by the support from the LHCb and ATLAS Gaudi
developers. Although not strictly their responsibility, they found the
time to offer help and support to this and the other adopting
experiments. Without this, the success of the adoption would have been
uncertain and at best would have taken much more effort. Daya Bay
recognized the need and importance of such support and, partly
selfishly, formed a mailing list~\cite{gaudi_db} and solicited the
involvement of Gaudi developers from many of the experiments involved
in its development and use. It became a forum that more efficiently
spread beneficial information from the main developers. It also
offloaded some support effort to the newly minted experts from the
other experiments so that they could help themselves.

There were, however areas that would improve the adoption of
Gaudi. While described specifically in terms of Gaudi they are general
in nature. The primary one would be direct guides on how to actually
adopt it. This is something that must come from the community and
likely in conjunction with some future adoption. Documentation on
Gaudi itself was also a problem particularly for Daya Bay developers
where many of the basic underlying framework concepts were new. Older
Gaudi design documents and some experiment-specific ones were
available but they were not always accurate nor focused on just what
was needed for adoption. Over time, Daya Bay produced its own Daya
Bay-specific documentation which unfortunately perpetuates this
problem.

Other aspects were beneficial to adoption. The Gaudi build system,
based on CMT~\cite{gaudi_cmt} is cross platform, open and easy to
port. It has layers of functionality (package build system, release
build system, support for experiment packages and ``external'' ones)
but it does not require a full all-or-nothing adoption. It supports a
staged adoption approach that allowed Daya Bay to get started using
the framework more quickly.

The importance of having all Gaudi source code open and available
cannot be diminished. Also important was that the Gaudi developers
included the growing community in the release process.

While Gaudi's CMT-based package and release build system ultimately
proved very useful, it hampered initial adoption as it was not
commonly used widely outside of Gaudi and the level of understanding
required was high. It is understood that there is now a movement to
provide a CMake based build system. This may alleviate this particular
hurdle for future adopters as CMake is widely used both inside and
outside HEP projects.

Finally, although Gaudi is full-featured and flexible it did not come
with all needed framework-level functionality and, in its core, does
not provide some generally useful modules that do exist in experiment
code repositories. In particular, Daya Bay adopted three Gaudi
extensions from LHCb's code base. These are actually very general
purpose but due to historical reasons were not provided
separately. These were GaudiObjDesc (data model definition), GiGa
(Geant4 interface) and DetDesc (detector description). Some extensions
developed by other experiments were rejected and in-house
implementations were developed. In particular, the extension that
provided for file I/O was considered too much effort to adopt. The
in-house implementation was simple, adequate but its performance was
somewhat lacking.

One aspect of the default Gaudi implementation that had to be modified
for use by Daya Bay was the event processing model. Unlike collider
experiments, Daya Bay necessarily had to deal with a non- sequential,
non-linear event stream. Multiple detectors at multiple sites produced
data in time order but not synchronously. Simulation and processing
did not preserve the same ``event'' multiplicity. Multiple sources of
events (many independent backgrounds in addition to signal) must be
properly mixed in time and at multiple stages in the processing
chain. Finally, delayed coincidence in time within one detector stream
and between those of different detectors had to be formed. The
flexibility of Gaudi allowed Daya Bay to extend its very event
processing model to add the support necessary for these features.

\medskip

\noindent{\bf\scshape{4.3~CMSSW and {\em art}}}\\
In  2005,   the  CMS  Experiment  developed   their  current  software
framework, CMSSW~\cite{cmssw},  as a replacement to  the previous ORCA
framework. The framework was built  around two guiding principles: the
modularity of  software development  and that exchange  of information
between  modules can  only  take place  through  data products.  Since
implementing  the  CMSSW, the  complexity  of  the CMS  reconstruction
software was  greatly reduced  compared with  ORCA and  the modularity
lowered the barrier  to entry for beginning  software developers. (The
Working Group thanks to Dr. Liz Sexton-Kennedy and Dr. Oli Gutsche for
useful discussions concerning the history and design of CMSSW.)

The CMS Software Framework is designed around four basic elements: the
framework, the event data model, software modules written by
physicists, and the services needed by those
modules~\cite{cmssw_source}. The framework is intended to be a
lightweight executable (cmsRun) that loads modules dynamically at run
time. The configuration file for cmsRun defines the modules that are
part of the processing and thus the loading of shared object libraries
containing definitions of the modules. It also defines the
configuration of modules parameters, the order of modules, filters,
the data to be processed, and the output of each path defined by
filters. The event data model (EDM) has several important properties:
events are trigger based, the EDM contains only C++ object containers
for all raw data and reconstruction objects, and it is directly
browsable within ROOT. It should be noticed that the CMSSW framework
is not limited to trigger based events, but this is the current
implementation for the CMS experiment. Another important feature of
the EDM over the ORCA data format was the requirement that all
information about an event is contained within a single file. However,
file parentage information is also kept so that if objects from an
input file are dropped (e.g., the raw data) that information can be
recovered by reading both the parent file and the current file in
downstream processes. The framework was also constructed such that the
EDM would contain all of the provenance information for all
reconstructed objects. Therefore, it would be possible to regenerate
and reproduce any processing output from the raw data given the file
produced from CMSSW. Another element of the framework that is useful
for reproducibility is the strict requirement that no module can
maintain state information about the processing, and all such
information must be contained within the base framework structures.

The {\em art} framework is an event processing framework that is an
evolution of the CMSSW framework.  In 2010, the Fermilab Scientific
Computing Division undertook the development of an experiment-agnostic
framework for use by smaller experiments that lacked the person-power
to develop a new framework. Working from the general CMSSW framework,
most of the design elements were maintained: lightweight framework
based on modular development, event data model, and services required
for modules. The output file is ROOT browsable and maintains the
strict provenance requirements of CMSSW. For Intensity and Cosmic
Frontier experiments, the strict definition of an event being trigger
based is not appropriate and so this structuring was removed and each
instance of {\em art} allows the experiment to define the event period of
interest as required. {\em art} is currently being used by the Muon g-2,
$\mu 2e$, NO$\nu$A, $\mu$BooNE, and LBNE/35T prototype experiments.

CMSSW did initially have some limitations when implemented, the most
significant being the use of non-interpreted, run-time configuration
files defined by the FHiCL language. The significance of this being
that configuration parameters could not be evaluated dynamically and
were required to be explicitly set in the input file. This limitation
meant it was impossible to include any scripting within the
configuration file. This limitation was recognized by the CMS
Collaboration and they quickly made the choice to instead transition
to Python (in 2006) based configuration files. At that time, a choice
was made that the Python evaluation of configuration code would be
distinctly delineated from framework and module processing. Therefore,
once the configuration file was interpreted, all configuration
information was cast as const within C++ objects and immutable. Due to
the requirement within CMSSW for strict inclusion of provenance
information in the EDM, the dynamic evaluation of configuration files
then cast as const parameters and stored in the EDM was not considered
a limitation to reproduction from raw data. When the {\em art} framework was
forked from CMSSW in 2010, the {\em art} framework reverted back to using
FHiCL language configuration files, and, while acceptable to
experiments at the time of adoption, some consider this a serious
limitation.

One of the challenges faced by the {\em art} framework has been the
portability of the framework to platforms other than Scientific Linux
Fermilab or Cern. The utilization of the Fermilab UPS and
cetbuildtools products within the build and release system that was
integrated into the {\em art} suite resulted in reliance upon those products
that is difficult to remove and therefore port to other platforms (OS
X, Ubuntu, etc). The CMSSW framework was implemented for CMS such that
the build system was completely available from source and mechanisms
for porting to experiment-supported platforms is integrated into the
build system. While portability of {\em art} is not an inherent problem of
the software framework design, and is currently being addressed by
both Fermilab SCD and collaborative experiments, it serves as a
significant design lesson when moving forward with {\em art} or designing
other frameworks in the future.

\medskip

\noindent{\bf\scshape{4.4~IceTray}}\\ 
IceTray~\cite{icetray} is the software framework used by the IceCube
experiment and also ported to SeaTray for the Antares experiment. The
framework is similar to other previously described frameworks in that
it takes advantage of modular design for development and
processing. Processing within the framework has both analysis modules
and services similar to those described for Gaudi, CMSSW, and {\em art}. The
IceTray framework and modules are written in the C++ language. The
data structure for IceTray is designated a ``frame'' and contains
information about geometry, calibration, detector status, and physics
events. Unlike other frameworks described, IceTray allows for multiple
frames to be active in a module at the same time.  This was
implemented due to the nature of the IceCube detector and the need to
delay processing an ``event'' until information from more than the
current frame is analyzed. This is accomplished through the use of a
uniquely designed I/O mechanism utilizing Inboxes and Outboxes for
modules. A module can have any number of Inboxes and Outboxes. The
development of IceTray was done within the IceCube experiment based
upon a specific set of requirements in 2003.

\medskip

\noindent{\bf\scshape{4.5~Opportunities for Improvement}}\\
Some best practices relevant to event processing frameworks are
identified: 

\medskip

\noindent{\bf open community}\\
Make source-code repositories, bug tickets and mailing lists (user and
developer) available for anonymous reading and lower the barrier for
accepting contributions from the community.  

\medskip

\noindent{\bf modularity}\\
Separate the framework code into modular compilation units with clear
interfaces which minimize recompilation. The system should work when
optional modules are omitted and allow different modules to be linked
at run-time.

\medskip

\noindent{\bf documentation}\\
Produce descriptions of the concepts, design and implementation of the
framework and guides on installation, extension and use of the
framework.

\medskip

The community should work towards making one event processing
framework which is general purpose enough to service multiple
experiments existing at different scales. This framework should be
ultimately developed by a core team with representation from multiple,
major stake-holder experiments and with an open user/developer
community that spans other experiments. Steps to reach this goal may
include:

\begin{itemize}
\item Form an expert working group to identify requirements and
  features needed by such a general use event processing
  framework. Much of this exists in internal and published notes and
  needs to be pulled together and made comprehensive.
\item The working group should evaluate existing frameworks with
  significant user base against these requirements and determine what
  deficiencies exist and the amount of effort required to correct
  them.
\item The working group should recommend one framework, existing or
  novel, to develop as a widely-used, community-supported project.
\item The working group should conclude by gauging interest in the
  community, survey experiments to determine what involvement and use
  can be expected and determine a list of candidate developers for the
  next step.
\item Assemble a core team to provide this development and support
  (something similar to the ROOT model). Direct support, which is
  independent from specific experiment funding, for some significant
  portion of this effort is recommended.
\end{itemize}

\medskip

\begin{center}
{\bf\scshape{5~~~Software Development}}
\end{center}

\noindent{\bf\scshape{5.1~Description}}\\
The tools supporting the full software development life-cycle can be
partitioned into these orthogonal categories.

\medskip

\noindent {\bf Code Repositories} store a historical record of
revisions to a code base including information on when a change is
made, the identity of the developer and some note explaining the
change. Repositories may be organized to hold a single logical unit of
source code (i.e., a {\em source package}) or may include multiple
such units relying on some internal demarcation. They allow diverging
lines of development, merging these lines and placing labels to
identify special points in the history (i.e., release tags).

\medskip

\noindent {\bf Package Build System} contains tools applied to the
files of {\em source package} in order to transform them into some
number of resulting files (executable programs, libraries). Typically
the system executes some number of commands (compilers, linkers) while
applying some number of build parameters (debug/optimized compilation,
locating dependencies, activating code features). This system may
directly install the results of the build to some area or in addition
it may collect the build results into one or more {\em binary
  packages}.

\medskip

\noindent{\bf Release Configuration} contains tools or specifications
for the collection of information needed to build a cohesive suite of
packages. It includes the list of packages making up the suite, their
versions, any build parameters, file system layout policy, source
locations, any local patch files and the collection of commands needed
to exercise the {\em package build system}.

\medskip

\noindent{\bf Release Build System} contains tools or processes
(instructions) that can apply a {\em release configuration} to each
{\em package build system} in the software suite. This process
typically iterates on the collection of packages in an order that
honors their inter-dependencies. As each package is built, the {\em
  release build system} assures it is done in a proper context
containing the build products of dependencies and ideally, controlling
for any files provided by the general operating system or user
environment. This system may directly install the results of the build
to some area and it may collect the build results into one or more
{\em binary packages}.

\medskip

\noindent{\bf Package Installation System} contains tools that, if
they are produced, can download and unpack {\em binary packages} into
an installation area. This system is typically tightly coupled to the
binary package format. It may rely on meta data internal or external
to the binary package file in order to properly resolve dependencies,
conflicts or perform pre- and post-installation procedures. The system
may require privileged access and a single rooted file system tree or
may be run as an unprivileged user and allow for multiple and even
interwoven file system trees.

\medskip

\noindent{\bf User Environment Management} contains tools that
aggregate a subset of installed software in such a way that the end
user may properly execute the programs it provides. This aggregation
is typically done through the setting of environment variables
interpreted by the shell, such as \texttt{PATH}. In other cases the
bulk of aggregation is done via the file system by copying or linking
files from some installation store into a more localized area and then
defining some minimal set of environment variables. In the case where
software is installed as system packages environment management may
not be required.

\medskip

\noindent{\bf Development Environment Management} contains tools to
assist the developer in modifying existing software or writing novel
packages. Such tools are not strictly required as a developer may use
tools from the above categories to produce a personal
release. However, in practice this makes the development cycle
(modify-build-test loop) unacceptably long. To reduce this time and
effort, existing release builds can be leveraged, installation steps
can be minimized or removed, and environment management can be such as
to use the build products in-place. Care is needed in designing such
tools to mitigate interference between individual developers while
allowing them to synchronize their development as needed.

\medskip 

\noindent{\bf Continuous Integration} contains tools and methodologies
for developing and exercising the code in order to validate changes,
find and fix problems quickly, and vet releases.

\medskip

\noindent{\bf Issues Tracker} contains tools to manage reporting,
understanding and addressing problems with the software, requests for
new features, and organizing and documenting releases.

\medskip

The following sections give commentary on what aspects are successful
for providing general, cross-experiment benefit and what failings are
identified. Explicit examples and areas where improvement may be made
are given.

\medskip

\noindent{\bf\scshape{5.2~Follow Free Software}}\\
The Free Software (FS) and the Open Source (OS) communities have a
large overlap with HEP in terms of how they develop and use
software. FS/OS has been very successful in achieving beneficial
sharing of software, largely due to that being a primary
goal of the community. It is natural then for the HEP software
community to try to emulate FS/OS.

Of course, the HEP community already benefits greatly from adopting
many libraries and tools from FS/OS. The community is relatively open
with its software development (in comparison to, for example,
industry).

There are however some ways in which the HEP community currently
differs from the FS/OS. Due to the nature of the HEP effort, some of
these differences are necessary, whereas in other areas improvements
can be made.

\begin{itemize}
\item Physics is the primary focus, not software. Of course this is
  proper. But, software is often not considered as important as other
  secondary aspects such as detector hardware design despite the fact
  that detector data is essentially useless today without quality
  software. Even in areas where software is the focus, often the ``hard
  core'' software issues are down-played or considered unimportant.
\item The use and development of HEP software is often tightly
  intertwined. End users of software are often its developers. Making
  formal releases is often seen as a hindrance or not performed due to
  lack of familiarity or access to easily usable release tools.
\item HEP software typically must be installed with no special
  permissions (non-``root''), in non-system locations, and with
  multiple versions of the software available on the same
  system. User/developers will often need to maintain locally modified
  copies of the software that override but otherwise rely on some
  centrally shared installation.
\item Versions matter a lot until they don't. A single code commit may
  radically change results and so upgrades must be done with care and
  changes constantly validated. Old versions must be kept accessible
  until new ones are vetted. They then become unimportant but must be
  forever reproducible in case some issue is found in the future which
  requires rerunning of the old version.
\item HEP software suites tend to be relatively large, often with the
  majority consisting of software authored by HEP physicists. Their
  design often requires an all-or-nothing adoption. Lack of careful
  modular components with well defined interfaces lead to design
  complexity and practical problems such as compilation
  time. Dependencies must be carefully handled and tested when
  lower-layer libraries are modified.
\end{itemize}

\medskip

\noindent{\bf\scshape{5.3~Category Integration}}\\
The categories described in Section 5.1 present some ideal
partitioning. Real world tools often cover multiple categories. When
this integration is done well it can be beneficial. Where it is not
done well it can lead to lock-in, lack of portability, increased
maintenance costs and other pathologies.

The functions of Configuration Management Tools (CMT) spans most of
these categories. Its design is such that it provides beneficial
integration with some capability to select the categories in which to
apply it. For example, it provides a package build system but one
which is flexible enough to either directly execute build commands or to
delegate to another package build system. This allows building and use
of external packages to achieve symmetry with packages developed by
the experiment. The configuration system is flexible enough to tightly
control versions for releases or to relax dependency conditions
suitable for a development context. The same information used to build
packages is used to generate shell commands to configure the end-user
environment.

CMT was initially used by LHC experiments but has successfully been
adopted by others outside of CERN (Daya Bay, Minerva, Fermi/GLAST, and
others). It is used across the three major computer platforms (Linux,
Mac OS X, and Windows).

In contrast is the UPS/CET system from Fermilab currently used to
build the {\em art} framework and its applications. UPS itself shares some
passing familiarity to CMT although its implementation is such that
even its proponents do not typically use it fully as it was
designed. Its entire ability to build packages is largely avoided. Its
other primary purpose of managing the user environment is often
augmented with custom shell scripts.

The CET portion adds a package build system based on CMake but with
hardwired entanglements with UPS. It tightly locks in to the source
code which versions of dependencies must be built against and the
mechanism to locate them. Developers commonly have their own effort
derailed if they attempt to incorporate work from others as any
intervening release forces their development base to become broken and
require reinitializing. Attempting to port the software ({\em art} and
LArSoft) entangled with this build system from the only supported
Linux distribution (Scientific Linux) to another (Debian) was found to
be effectively impossible in any reasonable time frame. This has led
to an effort by the LBNE collaboration to fully remove the UPS/CET
package build system from this software and replace it with one still
based on CMake but which follows standard forms and best practices. It
took far less effort to reimplement a build system than to adopt the
initial one. Effort is ongoing to incorporate these changes back into
the original software.

The astrophysics experiment LSST has developed a system,
EUPS~\cite{eups}, based on UPS, for code builds which allows local
builds on an experiment collaborators' laptop or server and which
probes the users local machine for already installed standard packages
(such as python). This system may be worth a look for smaller scale
experiments~\cite{lsst_wiki}.

\medskip

\noindent{\bf\scshape{5.4~Distributed Software Tools}}\\
Network technology has lead to paradigm shifts in binary code
distribution (e.g., CVMFS) and in distributing data (e.g.,
XRootD). HEP software development has always been very distributed and
it is important to continue to embrace this.

One successful embrace has been the move to \texttt{git} for managing
source code revisions. In comparison, any code development that is
still kept in Concurrent Versions System (CVS) or Subversion (SVN) is
at a relative disadvantage in terms of the ability to distribute
development effort and share its results.

Aggregating \texttt{git} repositories along with associated issue
trackers, web content (wikis) to provide a definitive, if only by
convention, center of development is also important. Some institutions
provide these aggregation services (Fermilab's Redmine) but the full
benefit comes when the software is exposed in a more global way such
as through online repository aggregators like GitHub or BitBucket.

Building software is an area that would benefit from a more
distributed approach. The traditional model is that the software
needed by the experiment is built from source by a site administrator
or an individual.  In some cases, an institution will take on the job
of building software for multiple experiments such as is done for some
experiments centered at CERN and Fermilab. While this service is
helpful for users of the platforms supported by the institution, it
tends to lock out users who do not use the officially supported
computer platforms. These unsupported platforms are otherwise suitable
for use and are often more advanced than the officially supported
ones. Small incompatibilities build up in the code base because they
go undetected in the relative monoculture created by limiting support
to a small number of platforms.

Distributed software build and installation systems are becoming
established and should be evaluated for adoption. Examples include the
package management systems found in the Nix and Guix operating
systems.  These allows one individual to build a package in such a way
that it may be used by anyone else. They also provide many innovative
methods for end-user package aggregation which leverage the file
system instead of polluting the user's environment variables.

Another example is Conda which provides a method to bundle up the
build configuration and a one-package unit of a release build
system. It also provides an end-user tool to install the packaged
results.  A coupled project is Binstar which can be thought of as a
mix between GitHub and the Python Package Index (PyPI). It allows
upload and distribution of packages built by Conda for later end-user
download and installation.

HEP community software projects and individual experiments can make
use of either the Nix/Guix or Conda/Binstar approaches to provide
ready-to-use code binaries to any networked computer in a trusted
manner. Sharing and coordinating the production of these packages
would take additional effort but this will be paid back by the
reduction of so much redundant effort that goes into building the same
package by each individual, experiment or project.

\medskip

\noindent{\bf\scshape{5.5~Automate}}\\
The majority of most HEP software suites are composed of four layers:
the experiment software on top is supported by general-use HEP
software. Below that is FS/OS packages which may be included in some
operating system distributions but in order to control versions and to
provide a uniform base for those OS distributions which do not provide
them, they are built from source. Finally, there is some lowest layer
provided by the OS. Each experiment draws each of these lines
differently and some may choose to blur them.

To produce proper and consistent release configurations and to track
them through time is challenging.  Once created, in principle, a
system can then apply these configurations in a way that automates the
production of the release build. Without this automation the amount of
effort balloons. This is only partially mitigated by distributing the
build results (addressed above).

Some experiments have developed their own build automation based on
scripts. These help the collaborators but they are not generally
useful.

CERN developed LCGCMT which, in part, provides automated building of
``externals'' via the LCG\_Builders component. This system is
specifically tied to CMT and is particularly helpful if CMT is adopted
as a release build system. This mechanism has been adopted by groups
outside of CERN, specifically those that also adopted the Gaudi event
processing framework. It has been specifically adopted by other
experiments.

Growing out of the experience with custom experiment-specific
automation and LCGCMT, the Worch~\cite{worch} project was developed to
provide build ``orchestration''. This includes a release configuration
method and an automated release build tool. It is extensible to
provide support for the other software development tool
categories. For example, it has support for producing needed
configuration files to provide support for using Environment Modules
as a method for end-user environment management.

\medskip

\noindent{\bf\scshape{5.6~Opportunities for Improvement}}\\
Some best practices in the area of software development tools are:

\medskip

\noindent{\bf Leverage Free Software} Rely on Free Software/Open
Source and do not become shackled to proprietary software or systems.

\medskip

\noindent{\bf Portability} Do not limit development to specific
platform or institutional details. 

\medskip

\noindent{\bf Automate} Produce builds and tests of software stacks in
an automated manner that is useful to both end-user/installers and
developers.

\medskip

Some concrete work that would generally benefit software development
efforts in HEP includes:

\begin{itemize}
\item Form a cross-experiment group to determine requirements for
  tools for build automation, software release management, binary
  packaging (including their format), end-user and developer
  environment management.
\item Form teams to design, identify or implement tools meeting these
  requirements. 
\item Assist experiments in the adoption of these tools.
\end{itemize}

\medskip

\begin{center}
{\bf\scshape{6~~~Data Management}}
\end{center}

\noindent{\bf\scshape{6.1~Definition}}\\
Data Management is any {\em content neutral} interaction with the
data, e.g., it is the data flow component of the larger domain of
Workflow Management (see Section 7.2.3). It addresses issues of data
storage and archival, mechanisms of data access and distribution and
curation -- over the full life cycle of the data. In order to remain
within the scope of this document we will concentrate on issues
related to data distribution, metadata and catalogs, and will not
cover issues of mass storage in much detail (which will be covered by
the Systems Working Group). Likewise, for the most part
network-specific issues fall outside of our purview.

\medskip

\noindent{\bf\scshape{6.2~Moving Data}}

\medskip

\noindent{\bf\em{6.2.1 Modes of Data Transfer and Access}}\\
In any distributed environment (and most HEP experiments are prime
examples of that) the data are typically stored at multiple locations,
for a variety of reasons, and over their lifetime undergo a series of
transmissions over networks, replications and/or deletions, with
attendant bookkeeping in appropriate catalogs. Data networks utilized
in research can span hundreds of Grid sites across multiple
continents.  In HEP, we observe a few different and distinct modes of
moving and accessing data (which, however, can be used in a
complementary fashion). Consider the following:

\medskip

\noindent{\bf Bulk Transfer}\\
In this conceptually simple case, data transport from point A to point
B is automated and augmented with redundancy and verification
mechanism so as to minimize chances of data loss. Such implementation
may be needed, for example, to transport ``precious'' data from the
detector to the point of permanent storage. Examples of this can be
found in SPADE (data transfer system used in Daya Bay) and components
of SAM~\cite{sam} and File Transfer Service at FNAL. Similar functionality
(as a part of a wider set) is implemented in the Globus Online
middleware kit~\cite{globus}.

\medskip

\noindent{\bf Managed Replication}\\
In many cases the data management strategy involves creating replicas
of certain segments of the data (datasets, blocks, etc.) at
participating Grid sites. Such distribution is done according to a
previously developed policy which may be based on storage capacities
of the sites, specific processing plans (cf. the concept of
subscription), resource quota and any number of other factors. Good
examples of this type of systems are found in ATLAS (Rucio) and CMS
(PhEDEx), among other experiments~\cite{rucio, phedex}.

\medskip

\noindent{\bf ``Data in the Grid'' (or Cloud)}\\
In addition to processing data which is local to the processing
element (i.e. local storage), such as a Worker Node on the Grid, it is
possible to access data over the net-work, provided there exists
enough bandwidth between the remote storage facility or device, and
the processing element. There are many ways to achieve this. Examples
include:

\begin{itemize}
\item using {\em http} to pull data from a remote location before
  executing the payload job. This can involve private data servers or
  public cloud facilities. 
\item utilizing XRootD~\cite{xrootd1, xrootd2} over WAN to federate
  storage resources and locate and deliver files transparently in a
  ``just-in-time'' manner. 
\item sharing data using middleware like Globus~\cite{globus}.
\end{itemize}

Choosing the right approaches and technologies is a two-tiered
process. First, one needs to identify the most relevant use cases and
match them to categories such as outlined above (e.g. replication vs
network data on demand). Second, within the chosen scenario, proper
solutions must be identified (and hopefully reused rather than
reimplemented).

\medskip

\noindent{\bf\em{6.2.2 From MONARC to a Flat Universe}}\\
The MONARC architecture is a useful case study, in part because it was
used in the LHC Run 1 data processing campaign, and also because it
motivated the evolution of approaches to data management which is
currently under way. It stands for {\em Models of Networked Analysis
  at Regional Centers}~\cite{monarc}. At the heart of MONARC is a
manifestly hierarchical organization of computing centers in terms of
data flow, storage and distribution policies defined based on
characteristics and goals of participating sites. The sites are
classified into ``Tiers'' according to the scale of their respective
resources and planned functionality, with ``Tier-0'' denomination
reserved for central facilities at CERN, ``Tier-1'' corresponding to
major regional centers while ``Tier-2'' describes sites of smaller
scale, to be configured and used mainly for analysis of the data (they
are also used to handle a large fraction of the Monte Carlo
simulations workload). Smaller installations and what is termed
``non-pledged resources'' belong to Tier-3 in this scheme, implying a
more ad hoc approach to data distribution and handling of the
computational workload on these sites. The topology of the data flow
among the Tiers can be described as somewhat similar to a Directed
Acyclic Graph (DAG), where data undergoes processing steps and is
distributed from Tier-0 to a number of Tier-1 facilities, and then on
to Tier-2 sites -- but Tiers of the same rank do not share data on a
peer-to-peer (P2P) basis. This architecture depends on the coordinated
operation of two major components:
\begin{itemize}
\item The Data Management System, that includes databases necessary to
  maintain records of the data parameters and location, and which is
  equipped with automated tools to move data between computing centers
  according to chosen data processing and analysis strategies and
  algorithms. An important component of the data handling is a
  subsystem for managing Metadata, i.e., information derived from the
  actual data which is used to locate specific data segments for
  processing based on certain selection criteria.
\item The Workload Management System (WMS) -- see Section 7 -- which
  distributes computational payload in accordance with optimal
  resource availability and various applicable policies. It typically
  also takes into account data locality in order to minimize network
  traffic and expedite execution. A mature and robust WMS also
  contains efficient and user-friendly monitoring capabilities, which
  allows its operators to monitor and troubleshoot workflows executed
  on the Grid.
\end{itemize}

While there were a variety of factors which motivated this
architecture, considerations of overall efficiency, given limits of
storage capacity and network throughput, were the primary drivers in
the MONARC model.  In particular, reconstruction, reprocessing and
some initial stages of data reduction are typically done at the sites
with ample storage capacity so as to avoid moving large amount of data
over the network. As such, it can be argued that the MONARC
architecture was ultimately influenced by certain assumptions about
bandwidth, performance and reliability of networks which some authors
now call ``pessimistic''~\cite{compmod} (p. 105).

At the time when LHC computing was becoming mature, great progress was
made in improving characteristics of the networks serving the LHC
projects. New generation of networks have lower latencies, lower cost
per unit of bandwidth and higher capacity. This applies to both local
and wide area networks~\cite{compmod} (p. 104).  This development
opens new and significant possibilities which were not available until
relatively recently; as stated in Ref.~\cite{compmod}:

\medskip

The performance of the network has allowed a more flexible model in
terms of data access:
\begin{itemize}
\item Removal of the strict hierarchy of data moving down the tiers,
  and allowing a more P2P data access policy (a site can obtain data
  from more or less any other site);
\item The introduction of the ability to have remote access to data,
  either in obtaining missing files needed by a job from over the WAN,
  or in some cases actually streaming data remotely to a job.
\end{itemize}

In practice, this new model results in a structure which is more
``flat'' and less hierarchical~\cite{compmod, lhccomp} and indeed
resembles the P2P architecture.

In principle, this updated architecture does not necessarily require
new networking and data transmission technologies when compared to
MONARC, as it mainly represents a different logic and policies for
distribution of, and access to data across multiple Grid sites. Still,
there are a number of differing protocols and systems which are more
conducive to implementing this approach than others, for a variety of
reasons:
\begin{itemize}
\item Reliance on proven, widely available and low-maintenance tools
  to actuate data transfer (e.g., utilizing HTTP/WebDAV).
\item Automation of data discovery in distributed storage.
\item Transparent and automated ``pull'' of required data to local
  storage. 
\end{itemize}

One outstanding example of leveraging the improved networking
technology is XRootD -- a system which facilitates federation of
widely distributed resources~\cite{xrootd3, xrootd4}. While its use in
HEP is widespread, two large-scale applications deserve a special
mention: it is employed in the form of CMS's ``Any Data, Anytime,
Anywhere'' (AAA) project and ATLAS's ``Federating ATLAS storage
systems using Xrootd'' (FAX) project, both of which rely on XRootD as
their underlying technology. ``These systems are already giving
experiments and individual users greater exibility in how and where
they run their workflows by making data more globally available for
access. Potential issues with bandwidth can be solved through
optimization and prioritization''~\cite{xrootd4}.

\medskip

\noindent{\bf\scshape{6.3~Metadata, Data Catalogs and Levels of Data
    Aggregation}}\\
To be able to locate, navigate and manage the data it has to be
described, or characterized. Metadata (data derived from the data) is
therefore a necessary part of data management. The list of possible
types of metadata is long. Some key ones are:
\begin{itemize}
\item Data Provenance: for raw data, this may include information on
  when and where it was taken. For processed data, it may specify
  which raw data were used. For many kinds of data, it is important to
  track information about calibrations used, etc.
\item Parentage and Production Information: one must keep track of
  software releases and its configuration details in each production
  step, be able to trace a piece of data to its origin (e.g., where it
  was produced, by which Task ID etc.), etc.
\item Physics: this may include analysis summary information or a
  specific feature characterizing a segment of data, e.g. type of
  events selected, from which trigger stream data was derived,
  detector configuration. 
\item Physical information: this might include the file size, check
  sum, file name, location, format, etc. 
\end{itemize}

Generally speaking, a data catalog combines a file catalog, i.e.,
information about where the data files are stored, with additional
metadata that may contain a number of attributes (physics, provenance,
etc.). This enables the construction of logical (virtual) data sets
like `WIMPcandidatesLoose' and makes it possible for users to select a
subset of the available data, and/or ``discover'' the presence and
locality of data which is of interest to the user. Grouping of data
into datasets and even larger aggregation units helps handle
complexity of processing which involved a very large number of
induvudual files. Here are some examples:

\medskip

\noindent{\bf Fermi Data Catalog}\\
Metadata can be created when a file is registered in the database. A
slightly different approach was chosen by the Fermi Space Telescope
Data Catalog. In addition to the initial metadata, it has a data
crawler that would go through all registered files and extract
metadata like number of events, etc. The advantage is that the set of
metadata then can be easily expanded after the fact by letting
loose the crawler with the list of new quantities to extract, which
then is just added to the existing list of metadata. Obviously this
only works for metadata included in the file and file headers.  Note
that since the Fermi Data Catalog is independent of any workflow
management system, any data processing metadata will have to be
explicitly added.

\medskip

\noindent{\bf SAM (Sequential Access Model)}\\
SAM is a data handling system developed at Fermilab. It is designed to
track locations of files and other file metadata. A small portion of
this metadata schema is reserved for SAM use and experiments may
extend it in order to store their quantities associated with any given
file.  SAM allows for the defining of a {\em dataset} as a query on
this file metadata. These datasets are then shorthand which can then
be expanded to provide input data to for experiment processes. Beyond
this role as a file catalog, SAM has additional functionality. It can
manage the storage of files and it can serve an extended role as part
of a workflow management system. It does this through a concept called
{\em projects} which are managed processes that may deliver files to
SAM for storage management and deliver files from storage elements to
managed processes. SAM maintains state information for files in active
projects to determine which files have been processed, which process
analyzed each file, and files consumed by failed processes. The
installation footprint required for SAM to be used at a participating
site depends on the functionality required. Lightweight access to
catalog functionality is provided via the SAM Web Services component
through a REST web interface which includes a Python client
module. Full features, including file management, requires a SAM
station installation and these exist at a small number of locations.

\medskip

\noindent{\bf ATLAS}\\
Distributed Data Management in ATLAS (often termed DDM) has always
been one of its focus areas, in part due to the sheer volume of data
being stored, shared and distributed worldwide (on multi-petabyte
scale), and to the importance of optimal data placement to ensure
efficiency and high throughput of processing~\cite{atlasdm}. Just like
with other major components of its systems, ATLAS has evolved its data
management infrastructure over the years. The system currently
utilized is Rucio~\cite{rucio}. We shall briefly consider basic
concepts and entities in this system pertaining to this section.

The atomic unit of data in ATLAS is the file. Above that, there are
levels of data aggregations, such as: 
\begin{itemize}
\item {\em dataset} Datasets are the operational unit of replication
  for DDM, i.e., they may be transferred to grid sites, whereas single
  files may not. Datasets in DDM may contain files that are in other
  datasets, i.e., datasets may overlap.
\item {\em container} Container is a collection of
  datasets. Containers are not units of replication, but allow large
  blocks of data, which could not be replicated to a single site, to
  be described in the system.
\end{itemize}

There are a few categories of metadata in Rucio:
\begin{itemize}
\item System-defined attributes (e.g., size, checksum, etc.)
\item Physics attributes (such as number of events)
\item Production attributes (parentage)
\item Data management attributes
\end{itemize}

\medskip

\noindent{\bf CMS}\\
CMS also employs the concept of a dataset. Metadata resides in, and is
being handled by the ``The Data Bookkeeping Service'' (DBS). This
service maintains information regarding the provenance, parentage,
physics attributes and other type of metadata necessary for efficient
processing. The Data Aggregation Service (DAS) is another critical
component of the CMS Data Management System. It ``provides the ability
to query CMS data-services via a uniform query language without
worrying about security policies and differences in underlying data
representations''~\cite{int_val}.

\medskip

\noindent{\bf\scshape{6.4~Small and Medium Scale Experiments}}\\
Small and medium scale research programs often have smaller needs
compared to the LHC or other large experiments. In these cases, it
will not be economical or feasible to deploy and operate the same kind
of middleware on the scale described in the previous Sections. Data is
often stored in a single or just a few geographical locations (`host
laboratories'), and data processing itself is less
distributed. However, many experiments today have data (or will have
data) characterized by volumes and complexity large enough to create
and demand a real data management system instead of resorting to
manual mode (files in unix directories and wiki pages). In fact, we
already find that some of the same elements, i.e., extensive metadata,
data catalogs, XRootD, are also used by some smaller experiments. The
main challenge here is the very limited technical expertise and
manpower available to develop, adapt and operate these sorts of tools.

With Cloud technology recently becoming more affordable, available and
transparent for use in a variety of applications, smaller scale
collaborations are making use of services such as Globus~\cite{globus}
to perform automated managed data transfers (cf. Section~6.2.1),
implement data sharing and realize the ``Data in the Cloud'' approach.
For small and mid-scale projects, platforms like Google Drive and
Dropbox offer attractive possibilities to share and store data at a
reasonable cost, without having to own much of the storage and
networking equipment and to deploy a complex middleware stack.

\medskip

\noindent{\bf\scshape{6.5~Opportunities For Improvement}}

\medskip

\noindent{\bf\em{6.5.1 Modularity}}\\
One problem with Data Management systems is that they often tend to
become monolithic as more and more functionality is added
(organically) -- see Section 7.1. While this may make it easier to
operate in the short term, it makes it more difficult to maintain over
the long term. In particular, it makes it difficult to react to
technical developments and update parts of the system. It is therefore
critical to make the system as modular as possible and avoid tight
couplings to either specific technologies or to other parts of the
ecosystem, in particular the coupling to the Workload Management
System. Modularity should therefore be part of the core design and
specifically separating the Metadata Catalogs from Data Movement
tools, with carefully designed object models and APIs. This would also
make reuse easier to achieve.

\medskip

\noindent{\bf\em{6.5.2 Smaller Projects}}\\
Smaller experiments have different problems. Most small experiments
have or will enter the petabyte era and can no longer use a manual data
management system built and operated by an army of graduate
students. They need modern data management tools. However, they have
neither the expertise to adapt LHC-scale tools for their use, neither
the technical manpower to operate them. Simplifying and downscaling
existing large scale tools to minimize necessary technical expertise
and manpower to operate them, even at the price of decreasing
functionality, may therefore be a good option.

A second option is to take existing medium-scale data handling tools
and repackage them for more general use. The problem is, however,
somewhat similar to what is described above. Often these systems have
become monolithic, have strong couplings to certain technologies and
significant work may be necessary to make them modular. This can be
difficult to achieve within the limited resources available and will
need dedicated support.

Finally, a few recent Cloud solutions have became available (and are
already used by small to medium size project), such as
Globus~\cite{globus}, Google Drive and Dropbox, among others. They do
provide a lot of the necessary functionality for data distribution and
sharing, and perhaps provide an optimal solution at this scale, when
combined with a flexible and reusable Metadata system (see notes on
modularity above).

\medskip

\noindent{\bf\em{6.5.3 Federation}}\\
Lastly, the success of Federated Storage built on XRootD shows the
importance of good building blocks and how they can be arranged into
larger successful systems.

\medskip

\begin{center}
{\bf\scshape{7~~~Workflow and Workload Management}}
\end{center}

\noindent{\bf\scshape{7.1~The Challenge of the Three Domains}}\\
In the past three decades, technological revolution in industry has
enabled and was paralleled by the growing complexity in the field of
scientific computing, where more and more sophisticated methods of data
processing and analysis were constantly introduced, oftentimes at a
dramatically increased scale. Processing power and storage were
becoming increasingly decentralized, leading to the need to manage
these distributed resources in an optimal manner. On the other hand,
increasing sophistication of scientific workflows created the need to
support these workflows in the new distributed computing medium. Rapid
evolution of the field and time pressures to deliver in this
competitive environment led to the design and implementation of
complete (to varying degrees) and successful solutions to satisfy the
needs of specific research projects. In general, this had two
consequences:
\begin{itemize}
\item Integrated and oftentimes -- not always -- project-specific
  design of workflow and workload management (see 7.2.3 for
  definitions).
\item Tight coupling of workflow and workload management to data
  handling components. 
\end{itemize}

We observe that there are essentially three interconnected domains
involved in this subject: Workflow Management, Workload Management,
and Data Management. In many systems (cf. Pegasus~\cite{pegasus})
some of these domains can be ``fused'' (e.g., Workflow + Workload). In
the following, we bring together a few standard definitions and real
life examples to help clarify relationships among these domains and in
doing so form the basis for possible HEP-FCE recommendations. Our goal
will be twofold:
\begin{itemize}
\item to identify the features and design considerations proven to be
  successful and which can serve as guidance going forward.
\item to identify common design and implementation elements and to
  develop understanding of how to enhance reusability of existing and
  future systems of this kind.
\end{itemize}

\medskip

\noindent{\bf\scshape{7.2~Description}}

\medskip

\noindent{\bf\em{7.2.1 Grid and Cloud Computing}}\\
According to a common definition, Grid Computing is the collection of
computer resources from multiple locations to reach a common
goal. Oftentimes additional characteristics added to this include
decentralized administration and management and adherence to open
standards. It was formulated as a concept in the early 1990s, and
motivated by the fact that computational tasks handled by large
research projects had reached the limits of scalability of most
individual computing sites. On the other hand, due to variations in
demand, some resources were underutilized at times. There was
therefore a benefit in implementing a federation of computing
resources, whereby large spikes in demand would be handled by
federated sites, while ``backfilling'' the available capacity with
lower priority tasks submitted by a larger community of
users. Technologies developed in the framework of Grid Computing (such
as a few reliable and popular types of Grid Middleware) became a major
enabling factor for many scientific collaborations including nuclear
and high-energy physics.

Cloud Computing is essentially an evolution of the Grid Computing
concept, with implied higher degree of computing resources and data
storage abstraction, connectivity and transparency of access. In
addition, Cloud Computing is characterized by widespread adoption of
{\em virtualization} -- which is also used in the traditional Grid
environment but on a somewhat smaller scale. At the time of writing,
``Cloud'' prominently figures in the context of commercial services
available on the Internet, whereby computing resources can be
``rented'' for a fee in the form of a Virtual Machine allocated to the
user, or a number of nodes in the Cloud can be dynamically assigned to
perform a neccesary computational task -- often as a transient,
ephemeral resource. Such a dynamic, on-demand characteristic of the
Cloud has led it to being described as an ``elastic'' resource. This
attribute is prominently featured in the name of the ``Amazon Elastic
Compute Cloud (EC2)''. This is an example of a {\em public} Cloud,
available to most entities in the open marketplace. Some organizations
choose to deploy Cloud technology on the computing resources directly
owned and controlled by them, which is then referred to as {\em
  private} Cloud.

Regardless of the obvious differentiation of Cloud computing (due to
its characteristics as a utility computing platform and pervasive
reliance on virtualization), many of its fundamental concepts and
challenges are common with its predesessor, Grid Computing. In fact,
the boundary is blurred even further by existing efforts to enhance
Grid middleware with tools based on virtualization and Cloud API which
essentially extend ``traditional'' Grid resources with on-demand,
elastic Cloud capability~\cite{grid_cloud}, leading to what is
essentially a hybrid system. The two are often seen as ``complementary
technologies that will coexist at different levels of resource
abstraction''~\cite{atlas_cloud}. Moreover, in parallel, many existing
grid sites have begun internal evaluations of cloud technologies (such
as Open Nebula or OpenStack) to reorganize the internal management of
their computing resources.  (See Ref.~\cite{atlas_cloud_2}).

In recent years, a few open-source, community developed and supported
Cloud Computing platforms have reached maturity, such as
OpenStack~\cite{openstack}. OpenStack includes a comprehensive set of
components such as ``Compute'', ``Networking'', ``Storage'' and others
and is designed to run on standard commodity hardware. It is deployed
at scale by large organizations and serves as foundation for
commercial Cloud services such as HP Helion~\cite{helion}. This
technology allows pooling of resources of both public and private
Clouds, resulting in the so-called {\em Hybrid Cloud}, where
technology aims to achieve the best characteristics of both private
and public clouds.

There are no conceptual or architectural barriers for running HEP-type
work ows in the Cloud, and in fact, efforts are well under way to
implement this approach~\cite{atlas_cloud}. However, there are caveats
such as 
\begin{itemize}
\item Careful overall cost analysis needs to be performed before each
  decision to deploy on the Cloud, as it is not universally cheaper
  than resources already deployed at research centers such as National
  Laboratories. At the same time, the Cloud is an efficient way to
  handle peak demand for the computing power due to its elasticity. It
  should also be noted that some national supercomputing centers have
  made part of their capacity available for more general high
  throughput use and may be an cost effective alternative.
\item Available bandwidth for staging in/staging out data in the Cloud
  (and again, its cost) need to be quantified and gauged against the
  project requirements.
\item Cloud storage cost may be an issue for experiments handling
  massive amounts of data~\cite{atlas_cloud} (p. 11)
\end{itemize}

\medskip

\noindent{\bf\em{7.2.2 From the Grid to Workload Management}}\\
Utilization of Grid sites via appropriate middleware does establish a
degree of resource federation, but it leaves it up to the user to
manage data movement and job submission to multiple sites, track job
status, handle failures and error conditions, aggregate bookkeeping
information and perform many other tasks. In the absence of
automation, this does not scale very well and limits the efficacy of
the overall system.

It was therefore inevitable that soon after the advent of reliable
Grid middleware, multiple physics experiments and other projects
started developing and deploying Workload Management Systems (WMS).
According to one definition, ``the purpose of the Workload Manager
Service (WMS) is to accept requests for job submission and management
coming from its clients and take the appropriate actions to satisfy
them''~\cite{egee}.  Thus, one of the principal functions of a Workload
Management System can be described as ``brokerage'', in the sense that
it matches resource requests to the actual distributed resources on
multiple sites. This matching process can include a variety of factors
such as access rules for users and groups, priorities set in the
system, or even data locality -- which is in fact an important and
interesting part of this process~\cite{panda}.

In practice, despite differing approaches and features, most existing
WMS appear to share certain primary goals, and provide solutions to
achieve these (to a varying degree). Some examples are:
\begin{itemize}
\item Insulation of the user from the complex and potentially
  heterogeneous environment of the Grid, and shielding the user from
  common failure modes of the Grid infrastructure.
\item Rationalization, bookkeeping and automation of software
  provisioning -- for example, distribution of binaries and
  configuration files to multiple sites.
\item Facilitation of basic monitoring functions, e.g., providing
  efficient ways to determine the status of jobs being submitted and
  executed.
\item Prioritization and load balancing across computing sites.
\item Implementation of interfaces to external data management
  systems, or actual data movement and monitoring functionality built
  into certain components of the WMS.
\end{itemize}

We shall present examples of existing WMS in one of the following
sections.

\medskip

\noindent{\bf\em{7.2.3 Workflow vs Workload}}\\
A scientific workflow system is a specialized case of a workflow
management system, in which computations and/or transformations and
exchange of data are performed according to a defined set of rules in
order to achieve an overall goal~\cite{swflow1, swflow2, pegasus}. In
the context of this document, this process involves execution on
distributed resources. Since the process is typically largely (or
completely) automated, it is often described as ``orchestration'' of
execution of multiple interdependent tasks. Work ow systems are
sometimes described using the concepts of a {\em control flow}, which
refers to the logic of execution, and {\em data flow}, which concerns
itself with the logic and rules of transmitting data. There are
various patterns identified in both control and
dataflow~\cite{wf_patterns}. A complete discussion of this subject is
beyond the scope of this document.

A simple and rather typical example of a workflow is often found in
Monte Carlo simulation studies performed in High Energy Physics and
related fields, where there is a chain of processing steps similar to
the pattern below:\\

\noindent{\em Event Generation $\Rightarrow$ Simulation $\Rightarrow$
  Digitization $\Rightarrow$ Reconstruction}\\

\noindent Patterns like this one may also include optional additional
steps (implied or made explicit in the logic of the workflow) such as
merging of units of data (e.g., files) for more efficient storage and
transmission.  Even the most trivial cases of processing, with one
step only, may involve multiple files in input and/or output streams,
which creates the need to manage this as a workflow. Oftentimes,
however, workflows that need to be created by researchers are quite
complex. At extreme scale, understanding the behavior of scientific
workflows becomes a challenge and an object of studies in its own
right~\cite{ext_wf}.

Many (but not all) workflows important in the context of this document
can be modeled as Directed Acyclic Graphs (DAG)~\cite{pegasus,
  task_atlas, swflow1}. Conceptually, this level of abstraction of the
workflow does not involve issues of resource provisioning and
utilization, monitoring, optimization, recovery from errors, as well
as a plethora of other items essential for efficient execution of
workflows in the distributed environment. These tasks are handled in
the context of {\em Workload Management} which we very briefly
described in Section 7.2.2.

In summary, we make a distinction between the Workflow Management
domain which concerns itself with controlling the scientific workflow,
and Workload Management which is a domain of resource provisioning,
allocation, execution control and monitoring of execution, etc. The
former is a level of abstraction above Workload Management, whereas
the latter is in turn a layer of abstractions above the distributed
execution environment such as the Grid or Cloud.

\medskip

\noindent{\bf\em{7.2.4 HPC vs HTC}}\\
The term High-Performance Computing (HPC) is used in reference to
systems of exceptionally high processing capacity (such as individual
supercomputers, which are typically highly parallel systems), in the
sense that they handle substantial workloads and deliver results over
a relatively short period of time. By contrast, in conventional usage,
HTC (High-Throughput Computing) involves harnessing a wider pool of
more conventional resources in order to deliver a considerable amount
of computational power, although potentially over longer periods of
time.  Note, however, that simulation campaigns, or long single runs
on HPC resources can often take as long as typical HTC
timescales. Nevertheless, it is reasonable to state that, ``HPC brings
enormous amounts of computing power to bear over relatively short
periods of time.  HTC employs large amounts of computing power for
very lengthy periods''~\cite{htc}.

In practice, the term HTC does cover most cases of Grid Computing
where remote resources are managed for the benefit of the end user and
are often made available on a prioritized and/or opportunistic basis
(e.g., the so-capped ``spare cycles'' or ``backfilling'', utilizing
temporary drops in resource utilization on certain sites to deploy
additional workload thus increasing the overall system throughput). A
majority of the computational tasks of the LHC experiments were
completed using standard off-the-shelf equipment rather than
supercomputers. It is important to note, however, that modern Workload
Management Systems can be adapted to deliver payload to HPC systems
such as Leadership Class Facilities in the US, and such efforts are
currently under way~\cite{hpc_wms, lecompte}.

\medskip

\noindent{\bf\em{7.2.5 The Role of Data Management}}\\
In most cases of interest to us, data management plays a crucial role
in reaching the scientific goals of an experiment. It is covered in
detail separately (see Section 6). As noted above, it represents the
dataflow component of the overall workflow management and therefore
needs to be addressed here as well.  In a nutshell, we can distinguish
between two different approaches to handling data -- on one hand,
managed replication and transfers to sites, and on the other hand,
network-centric methods of access to data repositories such as
XRootD~\cite{xrootd1, xrootd2}.  An important design characteristic
which varies widely among Workflow Management Systems is the degree of
coupling to the Data Management components. That has significant
impact on reusability of these systems as a more tight coupling
usually entails necessity of a larger and more complex stack of
software than would otherwise be optimal and has other consequences. 

\medskip

\noindent{\bf\scshape{7.3~Examples}}

\medskip

\noindent{\bf\em{7.3.1 The Scope of this Section}}\\
Workflow and Workload Management, especially taken in conjunction with
Data Management (areas with which they are typically interconnected)
is a vast subject and covering features of each example of WMS in any
detail would go well beyond the scope of this document. In the
following, we provide references to those systems which are more
immediately relevant to HEP and related fields than others.

\medskip

\noindent{\bf\em{7.3.2 HTCondor and glideinWMS}}\\
HTCondor~\cite{htcondor} is one of the best known and important set of
Grid and HTC systems. It provides an array of functionality, such as a
batch system solution for a computing cluster, remote submission to
Grid sites (via its Condor-G extension) and automated transfer
(stage-in and stage-out) of data. In the past decade, HTCondor was
augmented with a Workload Management System layer, known as
glideinWMS~\cite{giwms}. The latter abstracts remote resources (Worker
Nodes) residing on the Grid and effectively creates a local (in terms
of the interface) resource pool accessible by the user. Putting these
resources behind the standard HTCondor interface with its set of
utilities is highly beneficial to the users already having familiarity
with HTCondor since it greatly shortens the learning curve. On the
other hand, deployment of this system is not always trivial and
typically requires a central service to be operated with the desired
degree of service level (the so-called ``glidein factory'').

HTCondor has other notable features. One of the most basic parts of
its functionality is the ability to transfer data consumed and/or
produced by the payload job according to certain rules set by the
user. This works well when used in local cluster situations and is
somewhat less reliable when utilized at scale in the context of the
Grid environment. One of the HTCondor components, {\em DAGMan}, is a
meta-scheduler which uses DAGs (see 7.2.3) to manage
workflows~\cite{dagman}. In recent years, HTCondor has been augmented
with Cloud-based methodologies and protocols (cf. 7.2.1).

\medskip

\noindent{\bf\em{7.3.3 Workload Management for LHC Experiments}}\\
This is the list of systems (with a few references to bibliography)
utilized by the major LHC experiments - note that in each, we identify
components representing layers or subdomains such as Workload
Management etc.:

\medskip
\begin{center}
  \begin{tabular}{|c|c|c|c|} \hline \hline
    {Project} &{Workload Mgt} &{Workflow Mgt} &{Data Mgt}\\
    \hline
    ATLAS & PanDA~\cite{panda} & ProdSys2 & Rucio~\cite{rucio}\\
    CMS & GlideinWMS~\cite{giwms} & Crab3~\cite{crab3} &
    PhEDEx~\cite{phedex, phedex2}\\
    LHCb &  DIRAC~\cite{dirac} & DIRAC Production Mgt & DIRAC DMS\\
Alice & gLite WMS~\cite{glwms} & AliEn~\cite{alien} & AliEn~\cite{alien}\\
   \hline\hline
\end{tabular}
\end{center}

\medskip

\noindent{\bf\em{7.3.4 @HOME}}\\
There are outstanding examples of open source middleware system for
volunteer and grid computing, such as BOINC~\cite{boinc} (the original
platform behind SETI@HOME), FOLDING@HOME and
MILKYWAY@HOME~\cite{mwh}. The central part of their design is the
server-client architecture, where the clients can be running on a
variety of platforms, such as PCs and game consoles made available to
specific projects by volunteer owners. Deployment on a cluster or a
farm is also possible.

While this approach to distributed computing won't work well for most
experiments at the LHC scale (where moving significant amounts of data
presents a perennial problem) it is clearly of interest to smaller
scale projects with more modest I/O requirements. Distributed
platforms in this class have been deployed, validated and used at
scale.

\medskip

\noindent{\bf\em{7.3.5 European Middleware Initiative}}\\
The European Middleware Initiative~\cite{ems} is a consortium of Grid
services providers (such as ARC, dCache, gLite, and UNICORE). It plays
an important role in the the Worldwide LHC Computing Grid (WLCG).  The
{\em gLite}~\cite{glwms} middleware toolkit was used by LHC
experiments as one of the methods to achieve resource federation on
European Grids.

\medskip

\noindent{\bf\em{7.3.6 Fermi Space Telescope Workflow Engine}}\\
The Fermi workflow engine was originally developed to process data
from the Fermi Space Telescope on the SLAC batch farm. The goal was to
simplify and automate the bookkeeping for tens of thousands of daily
batch jobs with complicated dependencies all running on a general use
batch farm while minimizing the (distributed) manpower needed to
operate it. Since it is a general workflow engine it can be easily
extended to all types of processing including Monte Carlo simulation
and routine science jobs. It has been extended with more batch
interfaces and is routinely used to run jobs at IN2P3 Lyon, while
being controlled by the central installation at SLAC. It has also been
adapted to work in EXO and SuperCDMS.

\medskip

\noindent{\bf\scshape{7.4~Common Features}}

\medskip

\noindent{\bf\em{7.4.1 ``Pilots''}}\\
As mentioned in Section 7.2.2, one of the primary functions of a WMS
is to insulate the user from the heterogeneous and sometimes complex
Grid environment and certain failure modes inherent in it (e.g.,
misconfigured sites, transient connectivity problems, ``flaky'' worker
nodes, etc.). There is a proven solution to these issues, which
involved the so called late binding approach to the deployment of the
computational payload.

According to this concept, it is not the actual ``payload job'' that
is initially dispatched to a Worker Node residing in a remote data
center, but an intelligent ``wrapper'', sometimes termed a ``pilot
job'', which first validates the resource, its configuration and some
details of the environment (for example, outbound network connectivity
may be tested). In this context, ``binding'' means a matching process
whereby the payload job (such as a production job or a user analysis
job) which is submitted to the WMS and is awaiting execution is
assigned to a live pilot which has already perfomed validation and
configuration of the execution environment (for this reason, this
technique is sometimes referred to as ``just-in-time workload
management'').  Where and how exactly this matching process happens is
the subject of a design decision -- in PanDA, it is done by the
central server, whereas in DIRAC this process takes place on the
Worker Node by utilising the {\em JobAgent}~\cite{dirac2}.

With proper design, late binding brings about the following benefits:
\begin{itemize}
\item The part of the overall resource pool that exhibit problems
  prior to the actual job dispatch is excluded from the matching
  process by design. This eliminates a very large fraction of
  potential failures that the user would otherwise have to deal with
  and account for, since the resource pool exposed to the user (or for
  an automated client performing job submission) is effectively
  validated.
\item Some very useful diagnostic and logging capability may reside in
  the pilot. This is very important for troubleshooting and
  monitoring, which we shall discuss later. Problematic resources can
  be identified and flagged at both the site and worker node level.
\item In many cases, the overall latency of the system (in the sense
  of the time between the job submission by the user, and the start of
  actual execution) will be reduced -- due to the pilot waiting to
  accept a payload job -- leading to a more optimal user experience
  (again, cf. the ``just-in-time'' reference).
\end{itemize}

DIRAC was one of the first systems where this concept was proposed and
successfully implemented~\cite{dirac}.  This approach also forms the
architectural core of the PanDA WMS~\cite{pandawms}.

In distributed systems where the resources are highly opportunistic
and/or ephemeral, such as the volunteer computing we mentioned in
Section 7.3.4, this variant of the client-server model is the
essential centerpiece of the design. In BOINC, the ``core client''
(similar to a ``pilot'') performs functions such as maintaining
communications with the server, downloading the payload applications,
logging and others~\cite{boinc2}.

In HTCondor and GlideinWMS (see Section 7.3.2) there is no concept of
a sophisticated pilot job or core client, but there is a {\em glidein}
agent, which is deployed on Grid resources and which is a wrapper for
the HTCondor daemon process ({\em startd}). Once the latter is
initiated on the remote worker node it then joins the HTCondor
pool. At this point, matching of jobs submitted by the user to
HTCondor {\em slots} becomes possible~\cite{giwms}. While this ``lean
client'' provides less benefits than more complex ``pilots'', it also
belongs to the class of late-binding workload management systems,
although at a simpler level.

\medskip

\noindent{\bf\em{7.4.2 Monitoring}}\\
The ability of the user or operator of a WMS to gain immediate and
efficient access to information describing the status of jobs, tasks
(i.e. collections of jobs) and operations performed on the data is an
essential feature of a good Workload Management
System~\cite{panda2}. For operators and administrators, it provides
crucial debugging and troubleshooting capabilities. For the users and
production managers, it allows better diagnostics of application-level
issues and performance, and helps to better plan the user's
workflow~\cite{panda3}. All three domains involved in the present
discussion (Workflow, Workload, Data) benefit from a monitoring
capability.

\medskip

\noindent{\bf\em{7.4.3 The Back-End Database}}\\
The power of a successful WMS comes in part from its ability to
effectively manage state transitions of the many objects present in
the system (units of data, jobs, tasks, etc.). This is made possible
by utilizing a database to keep the states of these objects. Most
current implementations rely on a RDBMS for this function (e.g., ATLAS
PanDA is using the Oracle RDBMS at the time of writing). The database
serves both as the core intrument in the ``brokerage'' logic (matching
of workload to resources) and as the source of data for many kinds of
monitoring.

In seeking shared and reusable solutions, we would like to point out
that it is highly desirable to avoid coupling of the WMS application
code to a particular type or flavor of the database system, e.g., Oracle
vs. MySQL, etc. Such dependency may lead to difficulties in deployment
due to available expertise and maintenance policies at the target
organization and in some cases even licensing costs
(cf. Oracle). Solutions such as an ORM layer or other methods of
database abstraction should be utilized to make possible utilization
of a variety of products as the back-end DB solution for the Workload
Management System, without the need to rewrite any significant amount
of its code. This is all the more important in light of the widening
application of noSQL technologies in industry and research, since the
possibilities for future migration to a new type of DB must remain
open.

\medskip

\noindent{\bf\scshape{7.5~Intelligent Networks and Network
    Intelligence}}\\ 
Once again, the issue of network performance, monitoring and
application of such information to improve throughput and efficiency
of workflow belongs to two domains, Workload Management and Data
Management. In itself, the network performance monitoring is not a new
subject by any means and effective monitoring tools have been deployed
at scale~\cite{perfsonar}. However, until recently, network
performance data was not widely used as a crucial factor in managing
the workload distribution in HEP domain in ``near time''.  In this
approach, network performance information is aggregated from a few
sources, analyzed and used in determine a more optimal placement of
jobs relative to the data~\cite{panda4}.

A complementary strategy is application of ``Intelligent Networks'',
whereby data conduits of specified bandwidth are created as needed
using appropriate SDN software, and utilized for optimized data
delivery to the processing location according to the WMS logic.

\medskip

\noindent{\bf\scshape{7.6~Opportunities for Improvement}}

\medskip

\noindent{\bf\em{7.6.1 Focus Areas}}\\
Technical characteristics of Workload Management Systems currently
used in HEP and related fields are regarded as sufficient (including
scalability) to cover a wider range of applications and some existing
examples potentially support this point of view (cf. LSST and AMS
utilizing PanDA~\cite{panda3}). Therefore, the focus need not be on
entirely new solutions, but characterization of the existing systems
in terms of reusability, ease of deployment and maintenance, and
efficient interfaces. We further itemize this as follows:
\begin{itemize}
\item {\bf Modularity} -- addressing the issue of the ``Three Domains''
  (Section 7.1):
\begin{itemize}
\item {\bf WMS \& Data}: The interface of a Workload Management System
  to the Data Management System needs to be designed in a way that
  excludes tight coupling of one to another. This will allow
  deployment of an optimally scaled and efficient WMS in environments
  where a pre-existing data management system is in place, or where
  installation of a particular data management system puts too much
  strain on local resources. For example, replicating an instance of a
  high-performance and scalable system like Rucio which is currently
  used in ATLAS would be prohibitively expensive for a smaller
  research organization.
\item {\bf Workflow Management:} The concept of scientific workflow
  management is an old one, but recently it has come to the fore due
  to increased complexity of data transformations in many fields of
  research, in HEP and in several other disciplines. We recommend
  investigation of both existing and new solutions in this area, and
  design of proper interfaces between Workflow Management systems and
  underlying Workload Management systems.
\end{itemize}
\item {\bf Pilots:} The technique of deploying Pilot Jobs to worker
  nodes on the Grid adds robustness, flexibility and adaptability to
  the system. It proved very successful at extreme scale and the use
  of this technique should be encouraged. Creating
  application-agnostic templates of pilot code which can be reused by
  different experiments at scales smaller than LHC could be a
  cost-effective way to leverage this technique.
\item {\bf Monitoring}:
\begin{itemize}
\item {\bf Value}: A comprehensive and easy-to-use monitoring system
  has a large impact on the overall productivity of personnel
  operating a Workload Management System. This area deserves proper
  investment of effort.
\item {\bf Flexibility}: The requirements of experiments and other
  projects will vary, hence the need for flexible and configurable
  solutions for the Monitoring System utilized in Workload Management
  and Data Management. 
\end{itemize}
\item {\bf Back-End Database}: Ideally, there should be no coupling between
  the WMS application code and features of the specific database
  system utilized as its back-end -- this will hamper
  reusability. Such dependencies could be factored out and abstracted
  using techniques such as ORM, etc.
\item {\bf Networks}: There are significant efficiencies to be
  obtained by utilizing network performance data in the workload
  management process. Likewise, intelligent and configurable networks
  can provide optimal bandwidth for work ow execution.
\item {\bf Cloud}: Workload Management Systems deployed in this decade
  and beyond must be capable of efficiently operating in both Grid and
  Cloud environment, and in hybrid environments as well.
\end{itemize}

\medskip

\noindent{\bf\em{7.6.2 Summary of Recommendations}}

\medskip

\noindent{\bf WMS Inventory}\\
We recommend that future HEP planning activities should include an
assessment of major existing Workload Management Systems using
criteria outlined in Section~7.6.1, such as
\begin{itemize}
\item Modularity, which would ideally allow avoiding deployment of 
  monolithic solutions and would instead allow utilization of proper
  platform and technologies as needed, in the Data Management,
  Workload Management and Workflow Management domains.
\item Flexibility and functionality of monitoring.
\item Reduced or eliminated dependency on the type of the database
  system utilized as back-end. 
\item Transparency and ease of utilization of the Cloud resources.
\end{itemize}

Such an assessment will be useful in a number of complementary ways: 
\begin{itemize}
\item It will serve as a ``roadmap'' which will help organizations
  looking to adopt a system for processing their workflows in the Grid
  and Cloud environment, to make technology choices and avoid
  duplication of effort.
\item It will help identify areas where additional effort is needed to
  improve the existing systems in terms of reusability and ease of
  maintenance (e.g., implementing more modular design).
\item It will summarize best practices and experience to drive the
  development of the next generation of distributed computing systems.
\item It may help facilitate the developement of interopability layers
  in the existing WMS, which would allow future deployment to mix and
  match components from existing solutions in an optimal manner.
\end{itemize}

Further, this assessment should also contain a survey of existing
``external'' (i.e., open source and community-based) components that
can be utilized in existing and future systems, with proper interface
design. The goal of this part of the exercise it to identify cases
where software reuse may help to reduce development and maintenance
costs. For example, there are existing systems for flexible workflow
management which have not been extensively evaluated for use in HEP
and related fields.

It must be recognized that due to the complexity of the systems being
considered, the development of this assessment document will not be a
trivial task and will require appropriate allocation of
effort. However, due to the sheer scale of deployment of modern WMS,
and considerable cost of resources required for their operation, in
terms of both hardware and human capital, such an undertaking will be
well justified.

\medskip

\noindent{\bf Cloud Computing}\\
HEP experiments are entering the era of Cloud Computing. We recommend
continuation of efforts aimed at investigating and putting in practice
methods and tools to run scientific workflows on the Grid. Careful
cost/benefit analysis must be performed for relevant use cases.

In addition to extending existing WMS to the Cloud, we must work in
the opposite direction, i.e., to maintain efforts to evaluate
components of frameworks such as OpenStack~\cite{openstack} for
possible ``internal'' use in HEP systems.

\medskip

\begin{center}
{\bf\scshape{8~~~Geometry Information Management}}
\end{center}

\noindent{\bf\scshape{8.1 Description}}\\
Almost every HEP experiment requires some system for geometry
information management. Such systems are responsible for providing a
description of the experiment's detectors (and sometimes their
particle beam-lines), assigning versions to distinct descriptions,
tracking the use of these versions through processing and converting
between different representations of the description.

The geometry description is consumed mainly for the following
purposes: 
\begin{itemize}
\item Simulate the passage of particles through the detector or
  beamline.
\item Reconstruct the kinematic parameters and particle identity
  likely responsible for a given detector response (real or
  simulated).
\item Visualize the volumes to validate the description and in the
  context of viewing representations of detector response and
  simulation truth or reconstructed quantities.
\end{itemize}

The prevailing model for geometry information in HEP is Constructive
Solid Geometry (CSG). This model describes the arrangement of matter
by the placement of volumes into other volumes up to a top level
``world'' volume. It is typical to describe this as a daughter volume
being placed into a mother volume.  The placement is performed by
providing a transformation (translation plus rotation) between
conventional coordinate systems centered on each volume. A volume may
have an associated solid shape of some given dimensions and consist of
some material with bulk and surface properties. With no such
association the volume is considered an ``assembly'', merely
aggregating other volumes.

While this model is predominant, there is no accepted standard for
representing a description in this model. Instead, there are a variety
of applications, libraries and tools, each of which makes use of their
own in-memory, transient object or persistent file formats.

For example, Geant4 is a dominant particle-tracking Monte Carlo
simulation likely used by a majority of HEP experiments. It defines a
set of C++ classes for the description of a geometry and it can import
a representation of a description in the form of an XML file following
the GDML schema. It has various built-in visualization methods and can
export to some number of other formats for consumption by others.

Another common example are the \texttt{TGeo} classes provided by
ROOT. These can be constructed directly or via the import of a GDML
file (with some technical limitations). Like Geant4 objects, ROOT
provides means to track rays through the geometry as well as a few
visualization techniques.

There are stand-alone visualization tools such as HepRApp (which take
HEPREP files that Geant4 can produce), GraXML (which can read GDML
with some limitations or AGDD XML). There are also CAD programs that
can read OpenInventor files which can be produced. In experiments that
make use of Gaudi and DetDesc, the PANORAMIX OpenInventor based
visualization library can be used.

\medskip

\noindent{\bf\scshape{8.2 Unified System}}\\
This variety has lead to a ``tower of babel'' situation. In practice,
experiments limit themselves to some subset of tools. Developing their
own solutions is often seen as the least effort compared to adopting
others.  This of course leads to an ever larger tower. Some common
ground can be had by converting between different
representations. This is the approach taken by the Virtual Geometry
Model (VGM)~\cite{vgm} and the General Geometry Description
(GGD)~\cite{ggd}.

VGM provides a suite of libraries that allow for applications to be
developed which populate one system of geometry objects (e.g., Geant4
or ROOT). These can then be converted to a general representation and
finally converted to some end-form. Care must be taken to keep
implicit units correct and in an explicit system of units (the one
followed by Geant3). There are no facilities provided for the actual
authoring of the geometry description and that is left to the
application developer.

Addressing the need for an authoring system is a main goal of
GGD. This system provides a layered design. At the top is a simple
text-based configuration system which is what is expected to be
exposed to the end user. This drives a layer of Python builder modules
which interpret the configuration into a set of general in-memory,
transient objects. These objects all follow a strict CSG schema which
is conceptually compatible with that of Geant4. This schema includes
specifying an object's allowed quantities and provides a system of
units not tied to any particular ``client'' system. A final layer
exports these objects into other representations for consumption by
other applications, typically by writing a file. Access to both the
set of general geometry objects and any export-specific representation
is available for access in order to implement validation checks. Thus,
in GGD the source of geometry information for any ``world'' are the
Python builder modules and the end-user configuration files.

\medskip

\noindent{\bf\scshape{8.3 Problems with CAD}}\\
It is not unusual for an experiment at some point to consider
integrating CAD to their geometry information management system. CAD
provides far better authoring and visualization methods than most HEP
geometry software. It is typical that a CAD model for an experiment's
detectors or beamline must be produced for engineering purposes and it
is natural to want to leverage that information. However, there are
several major drawbacks to this approach.

CAD models sufficient for engineering purposes typically contain
excessive levels of information for HEP offline software
purposes. Applied to a tracking simulation this leads to an explosion
in the number of objects that must be queried on each step of the
particle's trajectory. It leads to large memory and CPU requirements
with minimal or incremental improvements in the quality of the
simulation or reconstruction results.

Use of CAD usually requires expensive, proprietary software with
significant expertise required to use.  This limits which individuals
can effectively work on the geometry and tends to lock in the success
of the experiment to one vendor's offering. It is typical for the
geometry description to require modification over a significant
portion of the experiment's lifetime and these limitations are not
acceptable.

Finally, the use of a CSG model is uncommon in CAD. Instead a
surface-oriented description is used. For its use in Geant4, a CSG
model is required. Converting from surfaces to CSG is very challenging
particularly if the CAD user has not attempted to follow the CSG model
in effect, if not in deed.

There is, however, potential in using CAD with HEP geometries. This is
being explored in GGD in the production of OpenInventor files which
can be loaded into FreeCAD, a Free Software CAD application.  While
FreeCAD can currently view an OpenInventor CSG representation it
cannot be used to produce them.  However, FreeCAD is extensible to new
representation types. With effort, it may be extended to produce and
operate on a suite of novel CSG objects which follow a schema
similarly to that required by Geant4.

\medskip

\noindent{\bf\scshape{8.4 Opportunities for Improvement}}\\
The ``tower of babel'' situation should be addressed by putting effort
in to following areas: 
\begin{itemize}
\item Form a small working group from geometry system experts to
  develop a formal data schema to describe the CSG objects that make
  up a general geometry system. This schema should be independent from
  any specific implementation but be consistent with major existing
  applications (specifically Geant4 and ROOT). The schema should be
  presented in a generic format but made available in a form that can
  be directly consumed (eg, JSON or XML) by software.
\item A general, transient data model for use in major programming
  languages (at least C++ and Python) should be developed which
  follows this schema. Independent and modular libraries that can
  convert between this data model and existing ones (GDML, ROOT)
  should be developed. One possibility is to further develop VGM in
  this direction.
\item Develop a general purpose geometry authoring system that can
  produce objects in this transient data model.
\end{itemize}

\medskip

\begin{center}
{\bf\scshape{9~~~Conditions Databases}}
\end{center}

\noindent{\bf\scshape{9.1 Description}}\\
Every HEP experiment has some form of ``conditions database''. The
purpose of such a database is to capture and store any information
that is needed in order to interpret or simulate events taken by the
DAQ system. The underlying principle behind such a database is that
the ``conditions'' at the time an event is acquired vary significantly
slower than the quantities read out by the DAQ in the event
itself. The period over which these condition quantities can change
range from seconds to the lifetime of the experiment.

In implementing a conditions database, an experiment is providing a
mechanism by which to associate an event to a set of conditions
without having to save a complete copy of those conditions with every
event. A secondary feature is that the event-to-conditions association
can normally be configured to select a particular version of the
conditions as knowledge about the conditions can change over time as
they are better understood.

\medskip

\noindent{\bf\scshape{9.2 Basic Concepts}}\\
It turns out that the basic concepts of a conditions database do not
vary between experiments. They all have the same issues to
solve. Questions of scale, distribution and so forth can depend on the
size and complexity of the data model used for quantities within the
database, but these aspects are secondary and are addressed by the
implementation. The resulting software for experiment differ more in
the choices of technologies used rather than in any conceptual
foundation.

\medskip

\noindent{\bf\em{9.2.1 Data Model}}\\
The Data Model of a conditions database defines how information is
grouped in atomic elements in that data base and how those atomic
elements are structure so that clients can recover the necessary
quantity. This is the most experiment specific concept as it is
directly related to the object model used in the analysis and
simulation codes. However the division of condition quantities into
the atomic elements is normally based on two criteria.
\begin{itemize}
\item The period over which a quantity varies, for example geometry
  may be updated once a year, while a detector calibration may be
  measured once a week.
\item The logical cohesiveness of the quantities, for example the
  calibrations for one detector will be separate from those of another
  detector even if they are updated at the same frequency.
\end{itemize}

\medskip 

\noindent{\bf\em{9.2.2 Interval of Validity}}\\
The standard way of matching a conditions element to an event is by
using a timestamp related to the event's acquisition. Given this time
the conditions database is searched for the instance of the element
that was valid at that time. (What to do when multiple instances are
valid for a given time is dealt with by versioning, see Section
9.2.3.) This therefore requires each entry in the conditions database
to have an interval of validity stating the beginning and end times,
with respect to the events timestamp, for which it should be
considered as the value for its quantity.

As analysis often proceeds sequentially with respect to events, most
implementations of condition database improve their efficiency by
caching the `current' instance of a quantity once it has been read
from the database until a request is made for a time outside its
interval of validity. At this point the instance appropriate to the
new time will be read in, along with its interval of validity.

\medskip

\noindent{\bf\em{9.2.3 Versioning}}\\
During the lifetime of an experiment a database will accumulate more
than one instance of a conditions element that are valid for a given
time. The two most obvious causes of this are the following.
\begin{itemize}
\item A conditions element is valid from a given time to the
  end-of-time in order to make sure there is always a valid instance
  of that element. At a later time during the experiment a new value
  for the element is measured and this is now entered into the
  database with its interval of validity starting later than the
  original instance but, as it in now the most appropriate from there
  on out, its validity runs until the end-of-time as well.
\item A conditions element may consist of a value derived from various
  measurements. In principle this can be considered a `cached' result
  of the derivation however it is treated as a first class element in
  the conditions database. At some point later, a better way of
  deriving the value is developed and this new value is placed in the
  database with the same interval of validity as the orignal one.
\end{itemize}

In both cases there needs to be a mechanism for arbitrating which
instances are used. This arbitration is managed by assigning versions
to each instance. The choice of which version to be used depends on the
purpose of the job that is being executed. If the purpose of the job is to use the `best' values then the `latest' version is used, but if the
purpose of the job is to recreate the results of a previous job or to
provide a known execution environment then it must be possible to
define a specific version to be used by a job.

In order for the above versioning to work there must be some monotonic
ordering of the versions. Typically this is done by the `insertion'
date which is the logical date when the version was added to the
database. It should be noted here that this date does not always
reflect the actual date the version was inserted, as that may not
create the correct ordering of versions.

\medskip

\noindent{\bf\scshape{9.3 Examples}}\\
The following is the subset of the implementations of a conditions
database pattern already done by HEP experiments.

\medskip

\noindent{\bf\em{9.3.1 DBI, Minos}}\\
This is a C++ binding that is decoupled from its surrounding framework,
a feature that allowed it be adopted by the Daya Bay Experiment for use
in its Gaudi customization. It has a very thin data model with a
conditions item being a single row in an appropriate SQL table.

\medskip 

\noindent{\bf\em{9.3.2 IOVSvc, Atlas}}\\
The Atlas Interval of Validity Service, IOVSvc, is tightly bound to
its Athena framework (their customization of the Gaudi framework). It
acts by registering a proxy that will be filled by the service rather
than direct calls to the service. It also has a feature where a
callback can be registered that is invoked by the service whenever the
currently value conditions item is no longer valid.

\medskip

\noindent{\bf\em{9.3.3 CDB, BaBar}}\\
The BaBar conditions database is notable in that during its lifetime
it went through a migration for an ObjectivityDB back-end to using
RootDB. This means that it can be used as an example of migrating
conditions database implementations as suggested in the next section.

\medskip

\noindent{\bf\scshape{9.4 Opportunities for Improvement}}\\
Given that the challenge of a conditions database, to match data to an
events timestamp, is universal and not specified to any style of
experiment, and given that numerous solutions to this challenge exist,
there is little point in creating new ones. The obvious opportunities
for improvement are ones that make the existing solutions available
for use by other experiment in order to avoid replication (again). To
this end the following approaches are recommended:
\begin{itemize}
\item A detailed survey of the interface of existing solutions should
  be made. The result of this survey should lead to the definition of
  a ``standard'' API for conditions databases that is decoupled for
  any particular framework.  This standard would be a superset of the
  features found in existing implementations so that all features of a
  given implementation can be accessed via this API. Suitable language
  bindings, such as C++ and python, so be provided as part of the
  standard.
\item Given the standard API and language bindings provided by the
  previous item, maintainers of conditions database implementations
  should first be encouraged, where possible, to develop the necessary
  code to provide the API as part of their implementation. They should
  then be encouraged to extend their own implementation to cover as
  much of the API as it is possible for their technology to support.
\item On the `consumer' side of the API, existing framework
  maintainers should be encouraged to adapt their framework so that it
  can use the standard API to resolve conditions database access. For
  frameworks and conditions databases that are tightly coupled, such
  as the Gaudi framework and its IOVsrc, this item, in concert with
  the previous one, will enable the conditions database to be
  decoupled for the analysis code.
\item In the longer term, given the standard conditions database API,
  the development of a Software-as-a-Service for conditions
  databases, for example using a RESTful interface, should be
  encouraged. This would allow the provisioning of conditions
  databases to be moved completely out of the physicist's realm and
  into that of computing support which is more suited to maintain that
  type of software.
\end{itemize}

\newpage

\begin{center}
{\bf\scshape{1~~~Systems Report: Introduction}}
\end{center}

The FCE Systems Working Group reviewed past and current practices and
future plans in the HEP computational program. The field is facing
increasing data rates, data volumes, and data processing needs in
existing and future experiments. The study drew from experience gained
in the field, documented in reports and presentations, and expert
testimony. Data-intensive research activities are a vital component in
HEP data analysis, and computing is a key enabler of these
activities. From several challenges and opportunities, two were
identified as being dominant:
\begin{itemize}
\item Data Storage and Data Access Technologies
\item Efficient Execution on Future Computer Architectures
\end{itemize}

The scope of the study covers the future path for computing systems
available across the HEP frontiers over the next decade. We address
advances in both the experiments and their data collection as well as
changes in computing architecture, networking, and I/O. The group
examined the previous decade, where and how success was achieved, in
addition to areas in which HEP did not take advantage of opportunities
available at the time, and which could have made dramatic differences
in how computing is used within HEP. Findings and recommendations were
guided by two major topics:
\begin{itemize}
\item Systems/Software related issues
\item The effects of changing technologies
\end{itemize}

\medskip

\begin{center}
{\bf\scshape{2~~~Computing Across HEP}}
\end{center}

In this Section we provide a broad overview of the computational
activities across the HEP landscape with a bias towards
systems-related issues. We cover the experimental domain in
Section~2.1 and the theoretical arena in Section~2.2.

\medskip

\noindent{\bf\scshape{2.1~HEP Experimental Workflows}}\\
Computational workflows in HEP experiments (and some involving theory
and modeling) are complex and the underlying infrastructure -- both
hardware and software -- takes a long time to plan and to implement,
requiring a sizable effort in manpower. We summarize the situation for
the three frontiers below, as well as provide a few abbreviated case
studies, DES and LSST for the Cosmic Frontier, and LHC for the Energy
Frontier. 

\medskip

\noindent{\bf\em{2.1.1 Cosmic Frontier}}\\
Computing for optical astronomy can be split into processing
images/spectra to produce measurements, and then analyzing the
measurements to reach a scientific conclusion. Typically, producing
measurements is a production activity, usually carried out by a
dedicated group.  For two Cosmology Frontier activities, DES and LSST,
this particular activity is primarily funded by NSF.  Unlike the
Intensity and Energy Frontier experiments, the process of extracting
measurements does not require simulation results.

The analysis of such data, however, is intimately tied to understanding
and constraining systematics through simulations, and thus is
considerably more data and compute intensive.  One method of analysis
is to compare data from observations to equivalent data derived from
cosmological simulations and modeling, expressing some candidate set of
physical properties.  Following from this, a conservative estimate of
the computing requirements set by current dark energy science is
significantly in excess of 500M core-hours.

Two representative examples of major Cosmic Frontier experiments are
the Dark Energy Survey (DES, ongoing) and the Large Synoptic Survey
Telescope (LSST, 2022). 

The software framework used for DES production is built on a grid
processing model, which allows DESDM to use remote computing sites
where appropriate.  DES data is transported nightly, as the data is
taken over routed research networks from Chile. The data is ingested
and nightly processing for supernova detection, as well as a first-cut
for initial image quality assessment is dispatched
automatically. Assessments of whether the data are of survey quality
are compiled, and inform the next night's observing program. A
different production cadence produces data for a release.  The release
production is organized into Final Cut, Multi-epoch, Mangle and PhotoZ
pipelines.  Nightly processing and release processing are supported by
the preliminary calibration and ``super'' calibration pipelines.

The production framework is such that a high level description of the
pipeline is produced. The description is parsed. Detailed lists of the
required files are generated.  Depending on the location of the bulk
computing site, computing resources are acquired, input files
transferred, and pipelines are started via Condor. The job progresses,
performing computations, and also uploading detected object catalogs
into an Oracle database at NCSA.  File-based data products are
returned to NCSA. Depending on the nature of the bulk computing site's
storage system, any, some, or all of the files associated with the job
may be left on the remote sites as a cache for future computations.

The recent revision of the production framework divides processing
into its natural independent units, such as exposure, co-add tiles and
SNE (supernova) fields. SNE files can now be dispatched on arrival;
exposures and co-additions can now fully occupy bulk computing
resources.  The system provides for unique file names; accurate
re-configurations; scalable meta-data collection; detailed file
provenance based on the Open Provenance Model; and detailed
operational and performance information. The system now supports
operation on an ensemble of computers, without requiring global file
systems (which have proved problematic for DES applications unless
very generously provisioned. This has enabled the use of Open Science
Grid resources on Fermigrid.
 
The DESDM production hardware infrastructure is centered at NCSA, but
includes important contributions for bulk computing from Fermilab and
NERSC.

DESDM software is organized into approximately 220 packages. DESDM
software is deployed in standalone tools and the above mentioned
pipelines. Each pipeline and tool has a configuration which is
logically independent of the others and must be specified
separately. A build system assures a consistent configuration of all
the dependent packages, builds and runs the provided unit tests for
each package, and makes packages available on the web. The
configuration manager installs the software on test and production
platforms. The installation procedure differs for each bulk computing
site: File system installs are required for iForge and the cosmology
cluster; Docker containers are required at NERSC and a CernVMFS
installation for FermiGrid.

The workflows are a series of stand-alone invocations of programs not
necessarily well-adapted for processing at the scale of DES.
Successive programs communicate with files.  The DES workflow
framework must deal with sequencing the program invocations, with naming
files appropriately, bringing more or fewer files back to support
debugging, extracting the relevant metadata and provenance records.
The DES workflows interact with an Oracle database at NCSA, recording
operational metadata, provenance and file metadata as workflows
progress.

All processing, even the local processing, uses the grid model.  The
community provided technologies supporting DES-developed production
framework software are:  Condor, DAGMAN, Condor glide-ins, Globus
Gatekeeper, and file transfers using http and gridFTP, and an Oracle
database.  DES used CernVMFS on Fermigrid and a batch system
integrated with Docker.  DES uses the EUPS sytem, which is also used
by LSST for configuration management of its software stacks, and has
integrated that system into its provenance model. Catalogs of detected
objects are views based on normalized relational tables. This
allows datasets to be updated incrementally. DES relies on the I/O
capability of Oracle, and its ability to be a parallel query engine.

In the case of LSST, cross-talk corrected data will be transferred
from the telescope site to a computing center in the AURA compound in
La Serena, Chile.  The data will be saved to disk, (and asynchronously
backed up to tape), and immediately forwarded to NCSA.  This action is
supported by redundant bandwidth-protected network paths that are
expected to achieve 40-100 Gpbs.

On arrival at NCSA, the data will be sent (asynchronously, again) to
tape, and immediately to a processing cluster, where alert production
occurs.  The goal of alert production is to detect and measure
transient objects (things which would be absent or would have moved if
the field was re-observed). An alert record will be assembled for each
transient object, which includes not only a detection announcement,
but also a cut out, orbit information, if relevant, etc.  No
``science'' is done -- what is produced are measurements allowing data
to be sorted by type by event brokers. On the cadence of an annual
release, data will be processed into release products.  There will be
pipelines that process the individual exposures. Data from the same
sky will be processed to increase depth.  Data re-assembled into
co-added exposures and support for science is not well served by
co-addition, such as the so-called ``multi-fit'' type techniques
needed, for example, by weak lensing. In the current baseline, data
will be released annually, except for year 1, which has two releases.
(There is a proposal that some of the processing take place at partner
institutions, such as IN2P3 in France). The LSST project will provide
10\% of its production capability to run user-provided codes via a
proposal/allocation process. Investigations are underway to determine
if it is reasonable to embed this processing into annual release
processing chains, as data movement would be reduced. Catalog data
will be ingested into an LSST developed relational database systems
named QSERV.  QSERV is a set of shared mySQL databases, with a
capability to batch table scans. Processed data will be released into
Data Access Centers, for distribution to those holding data rights.

\medskip

\noindent{\bf\em{2.1.2 Energy Frontier}}\\
Traditional High Energy Physics experiments record particle collisions
(events) with their detectors. Currently, the Energy Frontier is
represented by a single major experiment, the Large Hadron
Collider. The LHC has trigger rates of multiple hundreds of Hz and the
detectors have of the order of hundred million channels. In addition
to the raw data recorded by the detector, simulations of particle
collisions in the detectors are needed to extract physics results. The
experiment's application frameworks combine hundreds of software
components to simulate events and to reconstruct both simulated and
real detector collisions. Over the last two decades, hundreds of LHC
physicists and engineers have developed well over 10 million lines of
code, mainly in C++ and Python. The majority of this software is
experiment-specific, but several community projects such as ROOT,
Geant4, Gaudi, Frontier, and XRootD have emerged, and are now
well-established key components of the LHC software stack.

LHC applications store information in the form of objects, which are
persisted in files. Each file can contain different sets of objects of
the same event, grouped into data tiers. Major event contents are RAW
(raw) detector information, RECO information containing the
reconstructed detector signals and higher level event contents
optimized for analysis. Files are organized in datasets, grouping
events with similar physics and event content.

Several different types of workflows with different levels of
complexity are used to provide input to the physics analyses from
recorded and simulated collisions. Three major workflow types are data
reconstruction, user data analysis, and Monte Carlo
simulation. Traditionally, executables use a single core and have
access to 2 GB of memory, which represents a conventional limit for
the memory footprint of LHC applications. This limit became much more
of a concern after the migration from 32 bit to 64 bit
architectures. To fit within the 2 GB/core memory budget, LHC
experiments have invested heavily ($>10$ FTEs) in optimizing the
memory usage of their reconstruction software in particular, and still
have to split up their workflows in several pieces with intermediate
files written between the different stages of the workflows. These
intermediate outputs are usually transient but can also be permanently
stored to be reused.

Files are stored on mass storage systems with both disk replicas for
fast access and tape replicas for archival storage. Metadata is stored
in bookkeeping databases.

The LHC experiments rely on a distributed set of computing centers all
over the world, interconnected with strong networks. The initial
MONARC (Models of Networked Analysis at Regional Centers) model
expected the connectivity to be sufficient to exchange larger datasets
between clouds of sites hierarchically. But available network capacity
and reliability exceeded expectations so that a full mesh of
interconnectivity (P2P) could be established. This allowed for a
very flexible distribution of datasets for processing and analysis
access as well as opening up the possibility for direct WAN access to
files. The ability to access files through a capacious and reliable
network infrastructure allowed the collaborations to reduce the number
of dataset replicas initially needed by the LHC experiments to place
in close proximity to researchers over widely distributed regions
worldwide. Reducing dataset replicas allowed reducing the disk space
Tier-1 and Tier-2 centers had to provide, leading to significant
financial savings. Such changes in data replication policies and
modifications to the ATLAS and CMS data models have enabled the
collaborations to accommodate worldwide distributed analysis of vastly
increased datasets with only moderately increased storage resources.

Compute resources are accessible at the sites through batch systems
and GRID submission systems are used to access all resources in a
unified manner. Reliability and efficiency optimization caused the
change from a push model to a pull model based on initial pilot
submissions. For LHC run 2, more different resource types are
integrated into the same submission infrastructures such as commercial
and private clouds, supercomputing centers, direct access to batch
systems at university clusters, opportunistic resources on major GRID
infrastructures (EGI, OSG, NorduGrid, etc.).

Workflow management systems take care of using these submission
systems to split up work into manageable chunks (e.g., 8-hour jobs)
and to execute and monitor the jobs running on all the distributed
sites. Centrally managed production is only different from user-based
analysis by allowing users to write their own
non-experiment-sanctioned code and run it in the jobs. The entire
system requires a sustained multi-FTE effort to operate and optimize
the infrastructure as well as to include additional features.

\medskip

\noindent{\bf\em{2.1.3 Intensity Frontier}}\\
The Intensity Frontier represents a diverse set of high precision
experiments probing the limits of the Standard Model. Here, we will
focus on two types of experiments: Accelerator neutrino experiments
such as NO$\nu$A and those involving large liquid argon detectors and
precision muon experiments such as Mu2e and Muon g-2. Though the data
size and complexity of these experiments are not at the scale of CMS
or ATLAS, there are unique challenges that are pushing associated HEP
computing to the next level of size and complexity.

The accelerator neutrino experiments collect large amounts of data
with generally trigger-less or very loose trigger data acquisition
systems, since one correlates data in the detector with the
accelerator timing of the beam spill. Pattern matching and tracking
are the main challenges of these experiments, especially for those
with very large liquid argon TPCs. The tracking algorithms and
analyses to identify candidate tracks of one of the various neutrino
interactions or background processes are very complex and often
require large amounts of memory to execute. For example, the NO$\nu$A
experiment has a track identification algorithm that compares
candidate tracks in the data with 70 million simulated neutrino and
background tracks of various types. In order to execute this algorithm
with any speed, a special computer with 128 GB memory is used to load
the templates into memory and run the algorithm. The template library
needs to be about a factor of 10-100 larger to efficiently identify
tracks, but it is currently impossible to run the algorithm with a
library of this size. Algorithms and techniques from industry big
data, such as fast query NoSQL databases and specialized appliances
like the Cray Urika systems, are under investigation.

Precision muon experiments face different challenges. There, extreme
control of systematic uncertainties is absolutely paramount. Many
studies requiring large simulation runs are necessary to understand
such systematics from the apparatus and the data analyses. The main
challenge for these experiments is that they tend to be small. The
resulting difficulties are described below. 

For both neutrino and muon experiments, the physics models used in the
simulations are in an energy regime that has not been typical for past
and other experiments, such as those at the LHC. Large projects are
underway to determine and validate such models (e.g., the Genie
neutrino generator and physics models in Geant4).

\medskip

\noindent{\bf\em{2.1.4 Computational Issues for Smaller
    Experiments}}\\ 
Small experiments (e.g., $<200$ collaborators such as Muon g-2 and
Minerva) present unique problems in accomplishing the computing and
software necessary for the simulations, data taking and data
analyses. The main problem is a lack of personnel with extensive
experience in computing. Large experiments tend to have a sizable
cadre of senior staff, postdocs and students who form teams to write
low-level infrastructure and framework code, reconstruction code, the
simulation pipeline, data management systems, etc. Typically some
fraction of postdocs and students are actively engaged in these
computational and software activities. Such people simply do not exist on
the smaller experiments.

On small experiments there may be only one or two computing and
software savvy collaborators. Writing the many necessary systems from
scratch within the experiment is impossible. Furthermore, these
computing savvy people also have the responsibility to train incoming
(less experienced) postdocs and students who will write algorithms and
analysis code. The only solution is to use frameworks and systems from
elsewhere. For example, on Minerva, one of the main computing
collaborators brought the software framework from an experiment he
worked on in the past. But this solution still requires work, as such
systems are generally not written to be completely generic across
experiments and still must be maintained as new versions are
released. Furthermore, the original framework progresses and changes
without the adoptee experiment being kept in mind.

The other solution is to engage a larger, typically Lab-based, body of
computing expertise that writes framework and other code as a service
and also manages the computing hardware and infrastructure needed for
many experiments. Two good examples of generic frameworks written by
computing professionals and used by many experiments is the {\em art}
framework from the Fermilab Scientific Computing Division and the
FairROOT framework from GSI. Fermilab computing personnel also provide
data management and services to run jobs on the Grid. Furthermore,
they are forming a Production Operations Team of computing
professionals who will execute and monitor production simulation and
reconstruction workflows as a service for experiments. Educating new
users is also an important task of these large computing groups.

With these efforts, the experiments can concentrate on their specific
code and algorithms. This solution does raise interesting
challenges. There must be strong change-management processes to handle
feature requests to infrastructure code used by many experiments. Even
these Lab-based teams do not have large resources, so prioritizing the
needs of several experiments is challenging. Because these systems are
used for many different use cases and regimes, it is important to have
tests and excellent communication between the experiment's physicists
and the computing professionals.

\medskip

\noindent{\bf\scshape{2.2~Computing for HEP Theory}}\\
Major use of computing at HEP facilities and national HPC centers is a
characteristic of the HEP theory program. The part of the HEP
community that uses supercomputers at scale (accelerators, cosmology,
lattice QCD) also has strong interactions with vendors (e.g., IBM, Intel,
NVIDIA) to optimize code for next-generation systems as well as to
influence future computational architectures.

\medskip

\noindent{\bf\em{2.2.1 Cosmic Frontier}}\\
Interpreting future observations is impossible without a theory,
modeling, and simulation effort that matches the scale of current and
future sky surveys. To match the unprecedented precision and
resolution of the observations, required improvements over the next
decade are measured in orders of magnitude; new multi-physics and
multi-scale capabilities must be developed to address modeling of
complex physical processes. The flood of data from sky surveys has
dramatically reduced statistical uncertainties in cosmological
measurements and this trend will accelerate into the future. As a
result, large-scale theoretical modeling and data analysis are
required to open new discovery channels and to interpret results from
observations, such as statistical analyses of the galaxy distribution
across a large fraction of the observable sky. The role of computation
in what is now termed `precision cosmology' is thus pervasive,
complex, and crucial to the success of the entire enterprise.

Cosmological simulation codes such as HACC (collaboration led by
Argonne) and Nyx (collaboration led by LBNL) run on the largest
supercomputers available in the US. Large-scale simulation campaigns
can run up to hundreds of simulations and major individual simulations
can run at the full machine scale on leadership class systems (e.g.,
Mira at Argonne, Edison at NERSC, and Titan at Oak Ridge). Petabytes
of data can be produced by a single simulation.

As an example simulation framework, HACC (Hardware/Hybrid Accelerated
Cosmology Code) was developed originally for the heterogeneous
architecture of LANL's Roadrunner, the first computer to break the
petaflop barrier. HACC is designed with great flexibility in mind
(combining MPI with a variety of local programming models, e.g.,
OpenCL, OpenMP) and is easily adaptable to different platforms. HACC
is the first, and currently the only large-scale cosmology code suite
worldwide, that can run at scale (and beyond) on all available
supercomputer architectures. It was the first production science code
to run at greater than 10PFlops of sustained performance. Along with
HACC's simulation framework, a matched analysis system has been
co-developed for high-performance parallel in-situ and post-processing
analysis (statistical tools, halo and sub-halo finding, etc.). Some of
these algorithms have been embedded into ParaView, an open-source,
parallel visualization platform that has been recently enhanced for
visualization and analysis of cosmological simulations. LBNL and SLAC
researchers have developed an extensive set of wide-ranging simulation
analysis capabilities (the yt analysis code was incubated at SLAC and
work is ongoing at LBNL to merge yt and VisIt; yt can also be used as
a ParaView plugin); both groups have focused on tools for the
generation of synthetic sky catalogs for a number of cosmological
surveys.

Cosmological simulations use large-scale computing resources at DOE
supercomputer centers as well as offline analysis clusters. Storage is
typically at supercomputing sites, but limited local storage is also
available. Making large datasets and analysis capabilities available
to large collaborations remains a challenge.

\medskip

\noindent{\bf\em{2.2.2 Lattice QCD}}\\
Lattice QCD is an indispensable tool for HEP research, especially for
providing critical theoretical predictions and interpretations for
experimental programs.  Important applications include the QCD
thermodynamic, Quark-Gluon-Plasma, and Spin Physics of RHIC at BNL,
the hadronic contribution to the muon's anomalous magnetic moment
(g-2) at FNAL,  quark-flavor physics for B factories (Belle, BaBar,
and LHCb), and composite Higgs studies related to LHC physics by
solving strong dynamics. 

During the last decade, due to major improvements in theory,
algorithm, software and hardware, lattice QCD calculations have become
very accurate and reliable. Many of the basic quantities are now
computed with sub-percent total statistical and systematic errors.  It
has been important to have both in-house resources at the National
Laboratories, which support rapid developments of new ideas,
algorithms, and tuning of HPC code and parameters in a timely manner,
as well as access to leadership class computational resources for
production runs. The current HPC resources in U.S. Lattice QCD are
supported by the SciDAC program for software and DOE's LQCD project
for hardware, besides individual projects at the Laboratories. Support
for software development is distributed over the nation, and hardware
installed at BNL, FNAL, and J-Lab, is renewed periodically (roughly
every 5 years or so, jointly with the NP program).

The hardware requirement for lattice QCD is both in capability
computing (single stream of large jobs, needing a fast interconnect
and good parallel scalability) and in capacity computing (many streams
of intermediate size parallel jobs, typically with large memory and
I/O). The ensemble generation of QCD configurations is a typical
capability application, currently carried out on systems such as the
IBM Blue Gene/Q. On the BG/Q the largest job is run on 8 or 16 racks
(8,096 or 16,192 nodes), and achieved $1.6-3.2$~PFlops, or $30-40$\%
of the absolute peak speed with very good scaling. One stream of the
computation job typically lasts for a few months. The capacity machine
typically computes physical observables on the generated QCD ensemble,
and this task is performed on PC clusters or GPU clusters, currently
4K -- 8K core jobs for the CPU and up to 32 GPU parallel jobs with an
Infiniband interconnect. One capacity job duration is up to two weeks,
and the number of total jobs reaches to the tens of thousands.

The file size of the QCD configuration sample is of order 100~GB/file,
up to 10K files per parameter, with O(100) parameter points.  Beside
the QCD configuration sample, recent trends in I/O and disk usage for
the quark propagator and eigenvectors are such as to consume up to 100
TB/configuration. Typical temporary storage of O(100) configurations
means that 100~TB to 10~PB is typically used or even moved from one
site to another. For further details, the reader is directed to USQCD
white papers~\cite{usqcd}.

\medskip

\noindent{\bf\em{2.2.3 New Physics Searches and Perturbative QCD}}\\ 
In this subsection we follow the presentation given in
Ref.~\cite{exascale_wp_theory}.  Computational tasks in searches for
new physics are complicated by the large numbers of free parameters in
supersymmetric extensions of the Standard Model. Tools are available
to perform automated limit setting, but they require access to
reasonably large computational resources. Scans of parameter spaces
involve $\sim$250K-500K models, with simulation of LHC particle events
required for each individual model. For a single model set, typical
simulation datasets are of order $\sim$1-2~TB in size and require
$\sim$1-2M CPU-hours of computing time. Executables are single-core
for the most part and easily parallelized because of the underlying
Monte Carlo nature of the simulations. Memory limitations present in
new computational architecture do not present a bottleneck at the
current time.

High precision SM calculations rely on next-to-leading (NLO) QCD
perturbation theory results. Such calculations have played an
essential role in explicating the properties of the Higgs
boson. Automated NLO electroweak corrections will soon appear and
dedicated next-to-next-to leading order (NNLO) QCD calculations exist
for a number of important reactions. Precision calculations of this
type will become ever more important in the future as experimental
errors reduce and theory becomes more of a limiting factor. Current
NLO QCD calculations require 50-500K CPU-hours each, with storage
needs at the $\sim$1~TB scale. Parallelization strategies have included
both MPI and OpenMP (including hybrid approaches) and initial studies
for acelerated systems have been carried out. 

\medskip

\noindent{\bf\em{2.2.4 Accelerator Modeling and Simulations}}\\
Accelerator modeling makes extensive use of advanced computational
resources. The simulation tasks cover the domains of beam dynamics,
electromagnetics, and simulations for advanced accelerator
technologies. A given application is typically a challenging
multi-physics problem, often requiring large-scale parallel
resources. Until recently, the computational model has been a
distributed memory, MPI-based approach. Recently, however,
applications are moving to a hybrid MPI/OpenMP approach and GPU and
other acceleration techniques have been investigated. A recent
study~\cite{exascale_wp_acc} has concluded that future accelerator
modeling demands will be substantial, extending to scalable code
running at the million-core level.

\medskip

\begin{center}
{\bf\scshape{3~~~Software Development: Incompatibility with the
    Systems Roadmap}} 
\end{center}

The majority of HEP software, especially in the experimental arena,
has been architected in a way that does not immediately align with
ongoing and future developments in low-level computing
architectures. We discuss this issue with examples taken from the
different HEP frontiers.

\medskip

\noindent{\bf\scshape{3.1~Cosmic Frontier: DES}}\\
The DES pipelines are primarily constructed from project-specific
codes to remove instrumental signatures, and to identify and correct
instrumental defects. The high level operations that identify objects,
register the images on the sky and provide data products using input
data from many exposures are provided by a number of community codes.

In practice, this constrains production computing platforms to those
that are available and broadly accepted by community code
developers. The languages currently used and generally accepted by the
optical astronomy community are C, C++, and Python, along with
supporting libraries commonly used in astronomical image processing.
Only some programs are capable of parallelism or minimizing memory use
by windowing through an image.

Many workflows are executed on a single core, because not all steps in
a workflow can be parallelized. This is seen as an optimal strategy
for fully using computing, and produces high throughput, at the
expense of longer execution times. This is acceptable for many
production tasks. This serialization strategy, however is not an
optimal strategy for observing programs based on prompt transient
detection, which is of interest for other surveys.  We also recognize
that packing an independent workflow on one core is a strategy that
has a limited lifetime.

Given these constraints, the platforms used by DES are Intel and AMD
processors which use the X86 instruction set. Memory per core can be
as high as 8~GB. DES has experimented with ARM processors that are
emerging in the commodity server market.  DES cannot foresee the
community code base supporting GPUs in any way that makes assembling
the workflows tractable, and so has not investigated GPU
computing. The difficulty in moving to even an ARM platform is
identification and removal of low level software errors in codes that
are currently budgeted to be treat as black-boxes.

\medskip

\noindent{\bf\scshape{3.2~Energy Frontier: LHC Experiments}}\\
The LHC experiment frameworks are traditionally executed as
single-core applications. All rely on Intel-compatible
hardware. Optimizations for changes in clock speeds were performed by
changing the job splitting or defining different portions of work to
be done by a single application. (LHC reconstruction developers had to
face a ``memory crisis'' during Run 1.) Besides optimizing the
application software memory usage, LHC developers started exploring
new approaches to efficiently exploit the shared memory architectures
of multi-core CPUs. The first solution, multi-process event-parallel
frameworks, derived from the observation that, on Linux, a child
process will initially share all memory pages with the process it has
been forked from. Forking 8-event worker processes just before
entering the event-processing loop, a typical 8-core multi-process
reconstruction job uses roughly 75\% of the memory of 8 single-core
jobs.

A more radical solution, to which LHC experiments are migrating or
plan to migrate over the course of LHC Run 2, is to introduce
multi-threaded application frameworks that support sub-event parallel
processing. Besides using the shared memory of a current generation
multicore node even more efficiently (an 8-core multi-threaded
application uses roughly 60\% of the memory of 8 single-core
executables), a multi-threaded framework allows splitting the
reconstruction or simulation of an event into fine-grained tasks that
will be better suited to run in parallel on future many-core
architectures. One oft-mentioned issue with task-based parallelism is
that most LHC physicists and developers have not yet acquired the
skills to write code that can run efficiently, or at all, in a
multi-threaded environment. This will be addressed with education and
hands-on consulting with experts. Crucially, to support the gradual
migration of thousands of software components, Run 2 LHC frameworks
will have to allow existing components to run ``sandboxed'' in an
environment equivalent to a traditional single-thread application.

An added benefit of Run 2 memory-efficient concurrent frameworks is
that they will give application developers more flexibility in
deciding whether to optimize an algorithm or a data structure for
memory usage, CPU efficiency, or I/O speed.

\medskip

\noindent{\bf\scshape{3.3~Intensity Frontier}}\\
As computing architectures advance into new areas, such as
multicore/GPUs/Xeon Phi, there are systems and users that lag behind
the advances. In the Intensity Frontier, very few, if any, frameworks
take advantage of these new technologies. In many cases, the gains
realized by these technologies are not enough to justify the work
necessary to use them. For example, the Fermilab {\em art} framework can now
do event-level multiprocessing, but very few experiments are using
that feature. Larger gains in speed could be realized by fine grained
parallelism and utilization of co-processors/accelerators, but again,
converting those algorithms is a large task, and especially burdensome
for small experiments. Tracking for very large liquid argon detectors
may require some of these advanced technologies, and so LBNF/DUNE with
its large team of software-knowledgeable collaborators interested in
computing may have the resources to make progress.

\medskip

\noindent{\bf\scshape{3.4~Counter-Examples}}\\
In the case of theory and simulations, the employed codes are often
tuned to the latest generation of computational architectures. Lattice
QCD uses multiple CPU, and multiple/many cores with MPI and threading
(both in native pthreading or by OpenMP) effectively to the extent
possible, as well as GPU technology. Depending on the size of the
problem and other parameters, typically 10-40\% of the theoretical
peak speed is achieved in these codes, partly thanks to the intensive
efforts made possible by DOE SciDAC and other national/international
collaborations in the community. The current and near-future limiting
factor, however, is the small bandwidth of the inter-node
communication, which would have to be improved roughly by an order of
magnitude to reduce the imbalance between node-level flops and the
communication bandwidth. Unfortunately, the high cost of such an
improvement makes it unlikely to be implemented on large-scale
systems.

Similarly in the Cosmic Frontier, many years of effort have been spent
on maximizing the return from today's HPC resources to the fullest
extent possible. The Nyx code scales to ~100,000 cores on the
Leadership Class and NERSC computing facilities using both MPI and
OpenMP, while HACC has demonstrated similar scaling at the million
core level. In addition, HACC has been successfully optimized to work
in heterogeneous environments as well, and runs in full production
mode on CPU/GPU systems at very high levels of sustained performance.

\medskip

\begin{center}
{\bf\scshape{4~~~Effects of Changing Technologies}} 
\end{center}

In this section we review technologies that represent major building
blocks of the computing infrastructure used by HEP research
programs. For each of these technologies we provide a set of findings and
some recommendations.

\medskip

\noindent{\bf\scshape{4.1~Processors}}\\
A comprehensive treatment of the current evolution of computing
architecture can be found in the review by Kogge and
Resnick~\cite{kogge}. For roughly a decade, the failure of Dennard
scaling has driven chip design in new directions that emphasize
exploiting concurrency for gains in performance, while at the same
time introducing significant imbalances in floating-point performance
and off-chip communication bandwidth. Additionally, the amount of
DRAM/core is also reducing and the memory hierarchy is becoming more
complex (in-package memory, off-chip DRAM, off-chip
NVRAM). Historically, local data motion (data movement from memory to
CPU) has always been a key barrier in attaining optimal performance,
and this difficulty is now even more problematic. Finally, the
presence of multiple disparate architecture options (e.g., GPU, Xeon
Phi) raise new problems in the areas of programming models, tuning
algorithms to architectures, and portability.

GPU, many-core, and hybrid systems are likely to remain the workhorse
``beyond X86'' options for the next decade, although low-power
architectures and FPGA-based systems could also be interesting
alternatives. Other, radically different technologies are likely too
immature and narrow (e.g., neuromorphic computing) to provide
significant competition for general ``at scale'' applications on a
short timescale, but deserve attention. Even more forward-looking
technologies such as quantum computing have yet to overcome a number
of significant hurdles before they can transition from pure research
to production systems.


\medskip

\noindent{\bf\scshape{4.2~Software Challenges: Programmability
    versus Efficiency}}\\
The main challenge one faces in an environment of heterogeneity and
many-cores is that the scientist's ability to write a simple code in C,
C++, or Fortran will not take full advantage of the computational
power of a single node. The use of MPI, while advantageous for
internode communication and computation will, in general, hurt on-node
performance. Instead, programming models involving OpenMP, OpenCL,
CUDA, OpenACC (among others) will have to be utilized to get the most
out of the new systems. In addition, the programmer will have to be
aware of memory hierarchies, access speeds from the different cores to
this memory and more challenging NUMA effects that can become
bottlenecks to performance optimization.

Activities in which young researchers can get practical experience
programming on these next-generation heterogeneous machines, which
have appeared already at the LCF's at Argonne and Oak Ridge and will
come to NERSC in 2016, are critical as not only will these machines be
the backbone of the HPC world, but due to their overall cost and
energy efficiency, they will in fact begin to dominate the resources
available at universities and small clusters at Laboratories. An
increased relationship with vendors will also be potentially
beneficial as it will allow the pathway forward on the next-generation
chips to be created in an environment of co-design in which the needs
of the scientists have the potential of influencing certain design
choices.

Many activities required to move HEP computing in new directions need
investments in new ideas, novel algorithms, and optimization of
software and parameter tuning in a timely fashion before performing
large-scale production in the most efficient way.  For this reason,
local in-house resources at Laboratories and universities play an
important role, aside from access to larger-scale computational
facilities, e.g., national Leadership-class computational
resources.  Such a two-level resource model also provides a useful setting
to train young scientists.

\medskip

\noindent{\bf\scshape{4.3~Storage Hardware}}\\
Although magnetic disk recording was predicted to reach its limits
some years ago, developments in media and head technologies as well as
the increasing use of perpendicular recording have continued to push
the boundaries. It is fair to predict that magnetic disk technologies
will continue to result in increasing capacity and decreasing cost per
bit stored. The high rotational speed of magnetic disks, developed to
shorten data read time, would have shortened the life of drives
because of bearing failure, but the use of fluid suspension and gas
hydrodynamic bearings has vastly extended the life of the disk spindle
bearings.

``Cloud'' storage, though currently a hot topic, nonetheless relies on
conventional magnetic disk systems. To improve performance, some
large-scale servers use SRAM ``disks'' for buffer storage. Cloud
storage makes the details of the storage system transparent to the
user, though details of reliability and ownership of the data are
things the end user should not assume to be transparent. The costs of
transporting data to and from commercial clouds remain a concern.

Storage systems using RAID arrays are now standard, as are the
implementation of them in NAS and SAN storage systems. New are simpler
``just a bunch of disks'' (JBOD) and ``massive array of idle disks''
(MAID) systems. These are designed to be low cost and easy to
implement (JBOD) or to save power (MAID). The MAID systems have some
interesting cost statistics when the expenses for powering large disk
storage facilities and the needed air conditioning are included. As
the size of the ``server farms'' that HEP demands increases, power
saving ideas are likely to become more prominent.

Magnetic tape, whose demise was predicted decades ago, has also proven
to be resilient. Research has steadily increased the capacity of
magnetic tape systems and has kept the medium as the lowest cost per
bit stored of current media. It also shows a future growth path and
the LTO family of tape systems continues to be used for archival
storage, in particular. However, explicit archival disk storage based
on shingle recording methods, and with low power consumption, is being
announced in the market place. These systems may have a role for some
applications where tape is currently used.

Starting in 2015, DOE HPC systems are being procured with Burst
Buffers for additional storage and data processing capabilities. A
Burst Buffer is a combination of hardware and software meant to
leverage emerging NVRAM storage technologies in order to improve
application I/O effectiveness by simultaneously: reducing the time an
application spends doing I/O vs. computation; improving the efficiency
with which I/O is performed to underlying disk storage; and reducing
reads to the underlying disk filesystem by acting as a shared cache
for multiple processing elements. While this technology is in its
infancy, as it is more than an order of magnitude faster than spinning
disk, it has the potential to revolutionize the way HPC computing is
done. The underlying goal of the Burst Buffer is to provide a fast
storage system so as to improve overall application productivity and
resilience compared to a traditional filesystem. Although it is
predominately for checkpoint/restart, it is desirable for the Burst
Buffer subsystem to be a general-purpose solution for other
application needs, such as post-processing, in-transit visualization,
and data analytics.
  
\medskip

\noindent{\bf\scshape{4.4~Virtualization}}\\
One of the first implementations of massively distributed computing
for scientific research was the Grid. Grid computing essentially
combines computers from multiple administrative domains to solve
single, but independently parallelizable tasks. Compute provisioning
and management in grid computing is an extension and abstraction of
the traditional concept of batch computing, whereby individual
computing nodes are managed by a master system allocating tasks to the
nodes (i.e. workload management systems like PanDA and
glideinWMS). The distribution and allocation of computing and
data tasks across computing sites is done using specialized middleware
services responsible for different functions, such as authentication
and authorization, workload management, data transfers, logging, etc.

For workflows and provisioning computational resources, the term
virtualization is applied to techniques used to separate the work of
provisioning physical systems and operating system instances from the
view provided to applications.  The two main technologies are virtual
machines and container technology, such as Docker. A related
capability is the provisioning of ensembles of these machines along
with their associated storage.

As the use of virtualization has become a more and more viable and
efficient solution for instantiating computing nodes, the concept of
`the Cloud' or cloud computing has gradually established itself as a
more efficient and cost-effective solution for certain scientific
computing tasks. Although grid and cloud computing have many
similarities, cloud computing differs in a number of important aspects
for both providers and users. Compute provisioning and management in
cloud computing can make better use of virtualization and automation
thanks to an increasing number of standard tools and services, which
are supported both commercially and at a community level. This, in
turn, allows computing sites to provide increasing amounts of
resources, faster and more reliably. Cloud computing shifts the focus
from pure resource provisioning to service provisioning, allowing the
combination of different elements into higher level platforms or
applications, often tailored to specific user requirements.

This is an area being actively developed by industry, and as a result
there are a number of techniques, with more emerging.  There are many
complex drivers for the development of these techniques. These include
the need to achieve economies by sharing physical hardware, the need
to scale an application with varying demand, and the need to separate
the software delivery process from physical systems provisioning. Two
ways of deploying containers and virtual machine capabilities are
emerging for science use-cases.

One is to add these capabilities to traditional techniques in an
essentially conservative fashion. This involves integrating one or the
other basic virtualization techniques into existing clusters or HPC
systems.  An example of this work can be seen at NERSC, where batch
jobs can run code in Docker containers.  The existing batch system
serves as a resource manager, and cluster resources, such as task
allocation platform (TAP) systems, and global file systems are
imported into the containers. Expertise is thereby preserved.

The other way is essentially disruptive, with the physical resources and
containers being managed in new frameworks, of which the most notable
example available to the HEP community is OpenStack whose genesis is
cloud computing and is supported by industry-scale open source
projects. OpenStack provides very substantial capabilities, with a
corresponding complexity.

\medskip

\noindent{\bf\scshape{4.5 Networking}}\\
A key factor in better supporting HEP workflows lies in the
advancement of intelligent network services.  These services are
mostly an advancement of existing basic offerings or the development
of advanced capabilities that leverage disruptive technologies and/or
significant paradigm changes.

An emerging paradigm for next generation network architectures
revolves around the notion of the network as a ``multi-layer,
multi-technology'' construct over which multiple services can be
provided. These services include traditional IP routed services as
well as native access services for lower layers based upon Ethernet,
SONET/SDH, and wavelength-division multiplexing (WDM), and OpenFlow
technologies.  Making these services available via a well-defined
interface is critical in order to enable the next generation of
networked application innovations.  An expected shift to a more
interactive relationship between scientific applications and the
network has driven the creation of a new concept -- the ``Network
Service Plane'' (NSP).  The NSP can be broadly defined as a set of
abstracted network capabilities presented as provisionable service
objects that can interoperate and be incorporated into an application
resource provisioning workflow. The existence of this network service
plane enables the workflow management system to dynamically create and
manage a communication infrastructure to enable data analysis, rather
than be bound to constraints of deployed physical network connectivity
or long negotiations of new services from network service providers.

Described below are three major characteristics that are
fundamental to the next generation of network services.  The
combination of these characteristics is the enabler for a new class of
``Intelligent Network Services''.

{\em Network Services Interface:} A well-defined Network Service Interface
(NSI) which provides a distinct demarcation point between the network
and application/middleware must be developed.  This NSI should include
a set of atomic network services which are modular in nature and allow
for more complex services to be composed as needed.  This service
interface will need to include a new paradigm for how clients interact
with the network.  

{\em Network Service Capabilities Expansion:} Future network services must
expand in scope beyond current point-to-point Ethernet private
line services.  Features which are needed include multi-point
topologies, protection/restoration services, measurement services,
monitoring services, and security services. In addition, these
services will need to be provided across network infrastructures which
are heterogeneous in the technology and vendor dimensions.

{\em Scientific Workflow Support Features:} This is a set of capabilities
which allow for the network to add value and contribute to the
workflow, co-scheduling, and planning activities of application and
middleware systems.  The objective is to transform the network from a
passive participant (with little to no state awareness) to an active
participant in application/middleware workflow operations which are
responsive in ways that are meaningful to application/middleware
processes.  This requires the network (or agent on the networks'
behalf) to greatly increase its state awareness and intelligent
processing capabilities.

\medskip

\noindent{\bf\scshape{4.6 Non von Neumann Architectures}}\\
In addition to the changes we will see in mainline compute
architectures over the next decade, with many-core/multicore/GPU/etc.,
new technologies such as neuromorphic computing will become available
and can deliver huge benefits for certain types of data
analysis. Inspired by the human brain, neuromorphic chips are highly
suited for machine learning and pattern recognition. The neuromorphic
architecture consists of a scalable network of neurosynaptic cores
that can be programmed and trained. Algorithmically it can be
described as a neural network transforming a stream of input spikes
into a stream of output spikes. Late last year, IBM unveiled their
SyNAPSE/TrueNorth architecture and demonstrated that it can recognize
and classify objects with extremely low power
consumption~\cite{truenorth}.

The current production chip, the largest ever designed by IBM, has
4096 cores providing 1M neurons and 256M synapses and its designed to
be tiled in a 2D network. Due to its unique clock-less architecture,
the chip has a typical power density of 20mW/cm2, over 1000 times less
than a traditional CPU.  A single-rack system integrating 4096 chips,
the largest currently considered, would have 4B neurons, 1T synapses
($\sim1\%$ human brain) and consume $\sim$4~KW. Given the expected
data rates and signal time scales, a very promising application for
such a chip could be real-time data processing for the proposed LAr
TPC at E-LBNF and other Intensity Frontier experiments.

Current TPC track finding algorithms do not fully exploit the imaging
capabilities of a LAr TPC, discarding detailed signal shapes after
2D/3D-hit formation, with a potential loss of tracking efficiency
coming from ambiguities and tracks traveling along wire planes
\citep{usher}. A seedless, fully parallel, 4D track-finding algorithm
for LAR TPCs appears to be an ideal demonstrator of this new
architecture's pattern recognition capabilities. At this stage there
is no plan to produce a co-processor version of these chips, but there
are early ideas on how this could become part of an HPC
solution. Solutions like these would also be of interest for next
generation real-time pattern recognition triggers aimed at HL-LHC and
beyond.

\medskip

\begin{center}
{\bf\scshape{5~~~The HEP Distributed Computing Environment}} 
\end{center}

The geographically distributed infrastructure of HEP computing raises
important challenges, from managing worldwide distributed datasests
all the way down to provisioning and optimizing local computational,
network, and storage resources for a variety of HEP-specific
applications.

\medskip

\noindent{\bf\scshape{5.1~Resources and Resource Provisioning}}\\
As the computing environment becomes richer in the resources that are
expected to become available (e.g., combination of commercial and
institutional cloud services, traditional clusters, and a variety of
HPC systems), it is expected that, compared to the present situation,
computing in HEP will become significantly more complex. Managing
large-scale computational tasks will have to be carried out in an
environment where the diversity and the dynamic nature of the
computational resources will be essential control
variables. Statically federated resources will need to be integrated
with dynamically allocated resources causing new challenges for
resource planning, acquisition, and provisioning. A flexible
policy-based mechanism will be needed in order to comply with the ever
more complex user needs.

The enabling mechanism for satisfying the above requirements is to be
able to switch in the application and its software environment
seamlessly at execution time on any given resource, where the
individual resources must be capable of handling I/O dominated
computational traffic. Additionally, sufficiently fast, low-latency,
remote data transfer capabilities will be necessary. The overhead in
registering resources and adopting applications will need to be
minimized; a nominal requirement would be less than
10\% of the actual resource usage time.

\medskip

\noindent{\bf\scshape{5.2~HEP Applications and Networking}}\\
The distributed nature of HEP datasets and computational resources
places significant demands on the available networking resources. 

Because of the way applications make use of networks, certain key
opportunities exist for an application to change behavior based on
information observed about network characteristics and network
performance. These include the reliability and performance of
individual network connections (e.g., TCP connections between data
management systems), the reliability and performance of named entities
(e.g., the data management system at a particular research institution
considered as a whole), and the overall behavior of the entire system.

Since network performance is viewed primarily from the perspective of
the end systems, the information to be used in making performance
decisions is usually best collected by the end systems. There is an
exception to this, which is data from network measurement systems such
as perfSONAR, which collect performance data that can be used to
characterize portions of the network path. Therefore, applications
which make heavy use of the network would benefit both from tracking
their own metrics and from importing information from external network
measurement systems.

In order for next-generation advanced networked applications to be
successful, a set of network capabilities and services is needed, that
is significantly beyond what is available today.  This new class of
network must satisfy additional application-specific requirements to
feed the co-scheduling algorithms that will search for real-time and
scheduled resources, and will span the network and application
spaces associated with large volume, worldwide distributed data
analyses. Network infrastructures, capabilities, and service models
will need to evolve so that the network can be a key component of the
next technology innovation cycle.

The core requirements are as follows. Dynamic network service
topologies (overlays) with replication capabilities to support
reliable multicast are necessary to efficiently move the data from the
source to multiple depots.  (The conventional approach would be to
iteratively copy the data from the source to each depot individually.)
Resource management and optimization algorithms (e.g. available
network resources, load on target depot) are needed to effectively
determine the ``best'' depot to retrieve data from.

In order to meet these requirements a number of strategies will neeed
to be implemented. These include -- use of multi-constraint path finding
algorithms to optimize solutions (e.g. multicast for uploads, anycast
for downloads); scheduling and coordination of data replication in
phases to optimize the upload procedure (e.g., multicast source to
10~Gbps connected depots first, then use some of the depots as sources
to multicast data to 1~Gbps connected depots); downloading portions of
the files from various depots in parallel; network caching (store and
forward) to optimize transfers.

Aside from these more general requirements, there are some specific
situations that need to be discussed separately in the HEP
context. The first of these is the simultaneous use of multiple, very
large, distributed data sets via remote I/O. In some HEP workflows,
small portions of very large distributed data sets are needed for
analysis.  Accessing data subsets remotely is becoming practical as
remote file systems mature and as entire data sets become too large to
move. The requirements for doing this successfully include dynamic
network service topologies (overlays with multiple different paths)
with real-time networking for predictable network behavior since all
of the data sets will be accessed from a single running process;
co-scheduling of resources to access data sets at different locations;
data protection and/or recovery to prevent user processes from hanging
if remote data is inaccessible due to a network path going down; very
low packet loss and reordering may be necessary to prevent performance
collapse. In general, satisfying these conditions will require close
coupling or interaction between co-scheduled resources (e.g., network,
storage, compute) to build tolerance to service degradation and
macro-scheduling and coordination of remote file systems to optimize
read/write access across simultaneous distinct workflows.

Another important topic from the HEP perspective is the issue of
time-sensitive data transfers as part of execution workflows. Some
analysis-related workflows require geographically distributed
resources such as compute nodes, storage assets, and visualization
appliances to function as a single entity.  The seamless execution of
the workflow requires close coordination and co-scheduling of the
various components, including the network that ``glues'' the
components together, to ensure that the entire workflow pipeline is up
and functioning for tasks to run and complete in a timely manner. In
order for such distributed workflows to be practical, strict
co-scheduling of the necessary resources is necessary to ensure that
every component of the workflow pipeline is available and
connected. Management tools are needed that facilitate easy
composition of complex workflows and coordination of resources and
resource brokering facilities are required to expedite workflow
composition.

\medskip

\noindent{\bf\scshape{5.3~Global Data Access}}\\
This finding addresses issues observed with data access in HEP
collaborations that operate on a global scale. Thousands of scientists
at hundreds of institutions around the world are involved in data
analysis. Therefore, the problem arises of making the data, needed by
thousands of concurrent analysis processes, efficiently available,
independent of where the processes are running.

One can take a conservative approach by programmatically replicating
datasets to places that are in close proximity of the analysis
community, but this comes at the expense of having to provide the
storage resources for the same datasets multiple times, which, as the
LHC community has learned during Run~1, is not affordable. Or one can
take the approach of limiting the number of replicas to one or two and
use storage federations to discover the data inventory in real time
for direct access out of the analysis process. While the latter
approach is, resource-wise, a big advantage over multiple replicas, it
requires predictable high performance wide area networks to exist not
only within regions but also between them. Another approach is to let
the network ``learn'' what the data inventory is that the analysis
processes need in anticipation that there is overlap regarding what
analyzers are interested in in a particular region. Dynamic,
self-learning caches at major network exchange points could help
significantly to improve the data access efficiency associated with
processes accessing data over the wide area network directly. Several
technical approaches have been implemented to address issues of this
sort where the focus is on the needed data and not where it is
physically located. One example is the delivery of Web content, in
particular by the video streaming industry (e.g., Content Delivery
Networks), while others, e.g. Named Data Networks (NDN), are very
promising but still in the R\&D stage of development.

\medskip

\noindent{\bf\scshape{5.4~Systems Data Analytics}}\\
During the past several decades computing facilities contributing to
HEP computing as well as HEP scientific instruments have been
gathering not only enormous amounts of scientific data, but also very
large quantities of systems-monitoring data.  These data sets are well
known to be a valuable resource for designing and optimizing future
systems, since they provide access to actual resource utilization
patterns. Better curation, enrichment and management of this data
would improve its exploitation. Processing this data in near-real-time
may be useful in some cases.

The investigation of state-of-the-art data analytics in this sector
would be enhanced if the various groups concerned were to interact
more strongly.  In particular, a common data analytics platform might
be desirable (the ATLAS collaboration is working with CERN IT on
this). This area is somewhat recent in terms of R\&D activities within
HEP; a new ASCR/HEP SciDAC project has been initiated to exploit
systems data currently archived at Fermilab in order to optimize data
management and analysis.

\medskip

\noindent{\bf\scshape{5.5~Federated Identity Management}}\\
HEP researchers interact with people and resources domestically or
internationally, and they are thus often required to identify
themselves through some form of login. Unification of user
authentication is very important in enabling resource access in a
distributed complex computational environment. In particular,
satisfying security concerns of resource providers is an essential
requirement.

DOE Laboratories and universities have made some effort to have their
organization join InCommon (trust fabric for higher education and
research, operated by Internet2) but the process has not been
completed to the extent that the mechanism can be used smoothly and
throughout the community. It is very important that federated identity
management services are provided by organizations associated with the
HEP community. In addition, appropriate protocols need to be worked
out for automated HEP workflows to run on ASCR-controlled HPC
resources.

\newpage

\begin{center}
{\bf\scshape{Acronym Index}}
\end{center}

\noindent ACE3P --- Advanced Computational Electromagnetics 3-D
Parallel\\
\noindent AGDD --- ATLAS Generic Detector Description\\
\noindent AMR --- Adaptive Mesh Refinement\\
\noindent API --- Application Program Interface\\
\noindent ARM --- Advanced RISC Machines\\
\noindent ART --- Adaptive Refinement Tree\\
\noindent ASCR --- Advanced Scientific Computing Research\\
\noindent ATLAS --- A Toroidal LHC ApparatuS\\
\noindent AURA --- Association of Universities for Research in Astronomy\\
\noindent Belle II --- B detector at SuperKEKB\\
\noindent BLAST --- Berkeley Lab Accelerator Simulation Toolkit\\
\noindent BOINC --- Berkeley Open Infrastructure for Network
Computing\\
\noindent CAD --- Computer Aided Design\\
\noindent CAMPA --- Consortium of Advanced Modeling of Particle
Accelerators\\ 
\noindent CDB --- Conditions Database of BaBar\\
\noindent CHEP --- Computing in High Energy and Nuclear Physics\\
\noindent C-LIME --- (C-language)-Large Internet Message Encapsulation\\
\noindent CMS --- Compact Muon Solenoid\\
\noindent CMSSW --- CMS SoftWare\\
\noindent CMT --- Configuration Management Tools\\
\noindent CORSIKA --- COsmic Ray SImulations for KAscade\\
\noindent CP --- Charge Parity\\
\noindent CPS --- Columbia Physics System\\
\noindent CPU --- Central Processing Unit\\
\noindent CSG --- Constructive Solid Geometry\\
\noindent CUDA --- Compute Unified Device Architecture\\
\noindent CVS --- Concurrent Versions System\\
\noindent DAG --- Directed Acyclic Graph\\
\noindent DAGMAN --- Directed Acyclic Graph Manager\\
\noindent DAS --- Data Aggregation Service\\
\noindent DBS --- Data Bookkeeping Service\\
\noindent DES --- Dark Energy Survey\\
\noindent DESI --- Dark Energy Spectroscopic Instrument\\
\noindent DOE --- Department of Energy\\
\noindent DUNE --- Deep Underground Neutrino Experiment\\
\noindent EC2 --- Elastic Compute Cloud\\
\noindent EDM --- Event Data Model\\
\noindent EGI --- European Grid Infrastructure\\
\noindent EUPS --- Extended Unix Product System\\
\noindent EXO --- Enriched Xenon Observatory\\
\noindent GDML --- Geometry Description Markup Language\\
\noindent GGD --- General Geometry Description\\
\noindent GLAST --- Gamma-Ray Large Area Space Telescope\\
\noindent GPU --- Graphics Processing Unit\\
\noindent HACC --- Hardware/Hybrid Accelerated Cosmology Codes\\
\noindent HARP --- Hadron Production Experiment\\
\noindent HEP --- High Energy Physics\\
\noindent HEP-FCE --- High Energy Forum for Computational Excellence\\
\noindent HEPREP --- High Energy Physics REPresentables\\
\noindent HL-LHC --- High Luminosity Large Hadron Collider\\
\noindent HPC --- High Performance Computing\\
\noindent HTC --- High Throughput Computing\\
\noindent IMPACT --- Integrated Map and Particle Accelerator Tracking
code\\
\noindent IN2P3 --- National Institute of Nuclear Physics and Particle
Physics\\ 
\noindent LArTPC --- Liquid Argon Time Projection Chambers\\
\noindent LBNE -- Long-Baseline Neutrino Experiment\\
\noindent LBNF -- Long-Baseline Neutrino Facility\\
\noindent LCF --- Leadership Computing Facility\\
\noindent LCLS --- Linac Coherent Light Source\\
\noindent LHC --- Large Hadron Collider\\
\noindent LHCb -- Large Hadron Collider beauty experiment\\
\noindent IOVSvc --- Interval of Validity Service\\
\noindent JBOD -- Just a Bunch of Disks\\
\noindent JSON --- JavaScript Object Notation\\
\noindent LAACG --- Los Alamos Accelerator Code Group\\
\noindent LHC -- Large Hadron Collider\\
\noindent LHCb -- Large Hadron Collider Beauty Experiment\\
\noindent LSST --- Large Synoptic Survey Telescope\\
\noindent LTO --- Linear Tape-Open\\
\noindent MAID --- Massive Array of Idle Disks\\
\noindent MCNPX --- Monte Carlo N-Particle eXtended\\
\noindent MILC --- MIMD Lattice Computation\\
\noindent MONARC --- Models of Networked Analysis at Regional
Centers\\ 
\noindent MPI --- Message-Passing Interface\\
\noindent Mu2e --- Muon-to-Electron-Conversion Experiment\\
\noindent NAS --- Network-Attached Storage\\
\noindent NDN --- Named Data Networks\\
\noindent NERSC --- National Energy Research Scientific Computing
Center\\
\noindent NLO --- Next-to Leading\\
\noindent NNLO --- Next-to-Next-to Leading\\
\noindent NO$\nu$A --- NuMI Off-Axis $\nu$ Appearance\\
\noindent NSI --- Network Service Interface\\
\noindent NSP --- Network Service Plane\\
\noindent NUMA --- Non-Uniform Memory Access\\
\noindent NVRAM --- Non-Volatile Random-Access Memory\\
\noindent OpenACC --- Open Accelerators\\
\noindent OpenCL --- Open Computing Language\\
\noindent OpenMP --- Open Multi-Processing\\
\noindent ORM --- Object-Relational Mapping\\
\noindent OS --- Open Source\\
\noindent OSG --- Open Science Grid\\
\noindent P2P --- Peer to Peer\\
\noindent P5 --- Particle Physics Project Prioritization Panel\\
\noindent PanDA --- Production and Distributed Analysis System\\
\noindent PerfSONAR --- Performance  focused Service Oriented Network
monitoring ARchitecture\\ 
\noindent PhEDEx --- Physics Experiment Data Export \\
\noindent PM --- Particle-Mesh\\
\noindent P$^3$M --- Particle-Particle Particle-Mesh\\
\noindent PyPI --- Python Package Index\\
\noindent QCD --- Quantum Chromodynamics\\
\noindent RAID --- Redundant Array of Independ Disks\\
\noindent RDBMS --- Relational Database Management System\\
\noindent RHIC --- Relativistic Heavy Ion Collider\\
\noindent SAN --- Storage Area Network\\
\noindent SciDAC --- Scientific Discovery through Advanced Computing\\ 
\noindent SDH --- Synchronous Digital Hierarchy\\
\noindent SDN --- Software-Defined Networking\\
\noindent SONET --- Synchronous Optical Networking\\
\noindent SPH --- Smoothed-Particle Hydrodynamics\\
\noindent SQL --- Structured Query Language\\
\noindent SuperCDMS --- Super Cryogenic Dark Matter Search\\
\noindent SPH --- Smoothed Particle Hydrodynamics\\
\noindent TPC --- Time Projection Chamber\\
\noindent UNICORE --- Uniform Interface to Computing REsources\\
\noindent UPS --- Unix Product Support\\
\noindent VGM --- Virtual Geometry Model\\
\noindent WAN --- Wide Area Network\\
\noindent WebDAV --- Web Distributed Authoring and Versioning\\
\noindent WDM --- Wavelength-Division Multiplexing\\
\noindent WLCG --- Worldwide LHC Computing Grid\\
\noindent WMS --- Workload Management System\\
\noindent XML --- eXtensible Markup Language

\newpage

\begin{center}
{\bf\scshape{Acknowledgments}}
\end{center}

We record our thanks to our colleagues -- too many to name
individually -- for extensive discussions and for assisting with this
report in a number of places. In particular, we acknowledge the
significant community effort that went into the HEP Topical Panel
Meeting Report on Computing and Simulations in High Energy Physics and
in the Snowmass Community Summer Study. We have drawn extensively from
the data and insights presented in the associated reports in our
work. Partial support for this HEP-FCE activity was provided by DOE
HEP's Computational HEP program.

\newpage

\begin{center}
{\bf\scshape{Disclaimer}}
\end{center}

This document was prepared as an account of work sponsored by the
United States Government. While this document is believed to contain
correct information, neither the United States Government nor any
agency thereof, nor any of the universities associated with this
report, nor any of their employees, makes any warranty, express or
implied, or assumes any legal responsibility for the accuracy,
completeness, or usefulness of any information, apparatus, product, or
process disclosed, or represents that its use would not infringe
privately owned rights. Reference herein to any specific commercial
product, process, or service by its trade name, trademark,
manufacturer, or otherwise, does not necessarily constitute or imply
its endorsement, recommendation, or favoring by the United States
Government or any agency thereof, or by any of the universities
associated with this report. The views and opinions of authors
expressed herein do not necessarily state or reflect those of the
United States Government or any agency thereof or by any of the
universities associated with this report.

\end{document}